\documentclass[journal]{IEEEtran}
\usepackage{algorithm,algorithmic}
\usepackage[table]{xcolor}  
\usepackage{amssymb}
\usepackage{amsfonts}
\usepackage{graphicx}
\usepackage{pdfpages}
\usepackage{setspace}
\usepackage{multirow}
\usepackage{amsmath}
\usepackage{epsfig}
\usepackage{array}
\usepackage{color}
\usepackage{subfigure}
\usepackage{xcolor}
\usepackage{amssymb}
\usepackage{url}
\usepackage{mathrsfs}
\usepackage{booktabs}
\usepackage{cite}
\usepackage{bbding}
\usepackage{upgreek}
\usepackage{arydshln}
\usepackage{caption}
\usepackage{adjustbox}
\usepackage{makecell}
\usepackage{float}

\usepackage{subfig}
\newcommand{\blue}{\textcolor{blue}}

\ifCLASSINFOpdf
\else
\fi

\definecolor{chcol}{HTML}{EE0000}

\begin{document}
%
\title{\blue{A General Unfolding Speech Enhancement Method Motivated by Taylor's Theorem}}
%
%
%

\author{Andong~Li, \textit{Student Member, IEEE,}
		Guochen~Yu,
       Chengshi~Zheng, \textit{Senior Member, IEEE,}
       Wenzhe~Liu, and Xiaodong~Li
\thanks{Andong Li, Guochen Yu, Chengshi Zheng, and Xiaodong Li are with the Key Laboratory of Noise and Vibration Research, Institute of Acoustics, Chinese Academy of Sciences, Beijing, 100190, China, and also with University of Chinese Academy of Sciences, Beijing, 100049, China. (e-mail: {cszheng}@mail.ioa.ac.cn)}
\thanks{Wenzhe Liu is with Tencent Ethereal Audio Lab, 518000, Shenzhen, China. (e-mail:  wenzheliu@tencent.com)}
\thanks{Guochen Yu is also with State Key Laboratory of Media Convergence and Communication, Communication University of China, 100024, Beijing, China, and also with Key Laboratory of Media Audio and Video, Ministry of Education, Communication University of China, 100024, Beijing, China.}
\thanks{This work was supported in part by NSFC
(National Science Fund of China) under Grant No. 61571435. This work was also supported by the Open Research Project of the State Key Laboratory of Media Convergence and Communication, Communication University of China, China (No. SKLMCC2021KF014),}
\thanks{Manuscript received Nov. XX, 2021; revised XX XX, XXXX.}}

%
%

\markboth{Journal of \LaTeX\ Class Files,~Vol.~XX, No.~XX, December~XXXX}%
{Shell \MakeLowercase{\textit{et al.}}: Bare Demo of IEEEtran.cls for Journals}
%



\maketitle
\begin{abstract}
\blue{While deep neural networks have facilitated significant advancements in the field of speech enhancement, most existing methods are developed following either empirical or relatively blind criteria, lacking adequate guidelines in pipeline design.} Inspired by Taylor's theorem, we propose a general unfolding framework for both single- and multi-channel speech enhancement tasks. Concretely, we formulate the complex spectrum recovery into the spectral magnitude mapping in the neighborhood space of the noisy mixture, \blue{in which an unknown sparse term is introduced and applied for phase modification in advance}. Based on that, the mapping function is decomposed into the superimposition of the 0th-order and high-order polynomials in Taylor's series, where the former coarsely removes the interference in the magnitude domain and the latter progressively complements the remaining spectral detail in the complex spectrum domain. In addition, we study the relation between adjacent order terms and reveal that each high-order term can be recursively estimated with its lower-order term, and each high-order term is then proposed to evaluate using a surrogate function with trainable weights, so that the whole system can be trained in an end-to-end manner. \blue{Given that the proposed framework is devised based on Taylor's theorem, it possesses improved internal flexibility. Extensive experiments are conducted on WSJ0-SI84, DNS-Challenge, Voicebank+Demand, spatialized Librispeech, and L3DAS22 multi-channel speech enhancement challenge datasets. Quantitative results show that the proposed approach yields competitive performance over existing top-performing approaches in terms of multiple objective metrics.}
\end{abstract}

%
\IEEEpeerreviewmaketitle

\section{Introduction}
During the COVID-19 pandemic period, people tend to work at home and stay in connection with others relying on remote communication systems. In most cases, the recorded speech signals from the microphone are not clean and may contain various interferences such as environemental noises and reverberation components, which severely hamper the speech quality and intelligibility~{\cite{loizou2007speech}}, \blue{and also affect the performance of the back-end tasks such as automatic speech recognition (ASR)}. Speech enhancement (SE) is a typical ill-posed inverse problem and aims at removing the interference and preserving the target speech from  its noisy mixture~{\cite{loizou2007speech}}, which has become increasingly significant in modern electronic devices.

\blue{Over the past several decades, many traditional SE algorithms have been proposed.} Despite the effectiveness when specific statistical assumptions hold, these algorithms often suffer from heavy performance degradations under complicated acoustic conditions, \emph{e.g.}, low signal-to-noise ratios (SNRs) or highly non-stationary noise scenarios. Thanks to the proliferation of deep neural networks (DNNs), \blue{the speech enhancement task has been formulated into a supervised problem, and increasing attention has been attracted in recent years~{\cite{wang2018supervised}}}. Given a large set of synthesized noisy-clean pairs, the network can directly learn the non-linear relations from the noisy features to the clean target, which vastly simplifies the formulation procedure and gains overwhelming performance advantage over traditional methods.
\begin{figure}[t]
	\centering
	\centerline{\includegraphics[width=0.98\columnwidth]{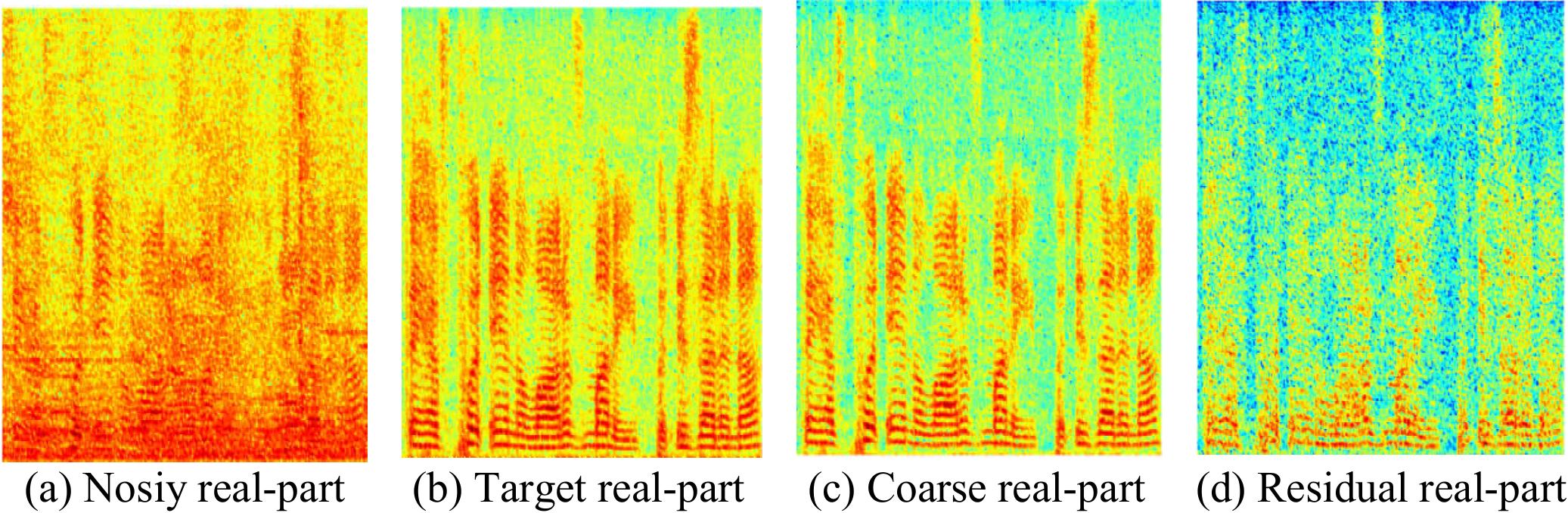}}
	\caption{Example illustration of the decoupled targets and only the real part is presented herein. (a) $\left| \mathbf{X} \right|\cos\left(\mathbf{\theta_{X}}\right)$: real part of the noisy spectrum. (b) $\left|\textbf{S}\right|\cos\left(\mathbf{\theta_{S}}\right)$: clean target spectrum. (c) $\left|\textbf{S}\right|\cos\left(\mathbf{\theta_{X}}\right)$: coarse estimated spectrum by coupling the clean magnitude with the noisy phase. (d) $\left|\textbf{S}\right|\left(\cos\left(\mathbf{\theta_{S}}\right) - \cos\left(\mathbf{\theta_{X}}\right)\right)$: residual component by subtracting (b) from (c).}
	\label{fig:example}
	\vspace{-0.65cm}
\end{figure}

In general, most of the previous NN-based approaches can be categorized into the time-frequency (T-F) domain~{\cite{tan2018gated, tan2019learning, hu2020dccrn, peer2022phase, li2021two, li2022filtering}} and time domain~{\cite{pascual2017segan, luo2019conv, luo2020dual, defossez2020real, pandey2020densely}}. While the short-time Fourier transform (STFT) is needed to convert waveform signals into the T-F spectrum for the former, both feature encoding and decoding operations are conducted in a trainable manner for the latter. In this study, we follow the first research direction as STFT provides a feasible and robust feature representation, where speech and noise components can exhibit clear structural characteristics in the spectral domain~{\cite{yin2020phasen}}. Earlier works usually estimate the spectral magnitude by either mapping~{\cite{xu2014regression, xiang2018speech}} or filtering~{\cite{hummersone2014ideal}} and leave the phase unaltered, which inevitably restricts the upper-bound performance. More recently, the importance of phase recovery in subjective perception has been confirmed~{\cite{paliwal2011importance}}, and a flurry of research methods incorporating phase were proposed. \blue{Common estimation strategies for phase-aware SE include complex ratio mask (CRM)~{\cite{williamson2015complex}} or complex spectral mapping (CSM)~{\cite{tan2019learning}}.} In~{\cite{9552504}}, the compensation effect was formally revealed in the complex-domain spectrum optimization, where the magnitude inclines to get distorted to compensate for the phase accuracy~{\cite{xiaoxue}}. To alleviate this issue, more recently, a two-stage decoupling-style estimation strategy was proposed in~{\cite{li2021two}}. To be specific, \blue{the magnitude estimation is first considered as the prior information}. Then, the residual learning is introduced, \emph{i.e.}, the complex-valued residual component is estimated with a sparse distribution, which measures the phase gap between the noisy and clean spectra. By doing so, we decouple the magnitude and phase optimization targets in a step-wise manner. Afterward, in~{\cite{li2022filtering}}, a collaborative estimation method was adopted to implement magnitude filtering and residual estimation in parallel, which improves the overall inference efficiency. An example of the decoupled targets is illustrated in Fig.~{\ref{fig:example}}.

\blue{Although the efficacy of the decoupling-style estimation scheme has been demonstrated beyond doubt, its reliance on empirical methods and lack of a theoretical foundation may pose limitations. Upon revisiting previous neural network-based works, it is observed that many state-of-the-art (SOTA) models focus primarily on maximizing performance by introducing complex network modules~{\cite{yu2022dbt, yin2022tridentse}} or using parallel and/or cascade multiple processing stages~{\cite{wang2021neural,zhongqiu2022l3das22,zhongqiu2022lowdistortion}}, without adequately considering the feasibility of the model design. Such an approach may result in models lacking guidelines and reduced flexibility.} \blue{To this end, in this paper, we revisit the magnitude-phase decoupling operation and convert the previously proposed ``filtering-and-refining'' strategy in~{\cite{li2022filtering}} into a more generalized format, \emph{i.e.}, ``filtering-and-progressively-refining''.} Specifically, we formulate the original complex spectrum recovery problem into the function approximation in the neighborhood space of the input spectrum. In other words, \textit{\blue{if the unknown complex-valued sparse term is assumed to exist} and can be utilized to recover the phase in advance, the spectrum can then be perfectly recovered in theory by magnitude-only estimation}, \emph{i.e.}, $S = \mathcal{G}\left(X+\delta\right)$, \blue{where $\left\{S, X, \delta \right\}$ respectively denote the clean, noisy and unknown sparse terms}, and $\mathcal{G}\left(\cdot\right)$ denotes the magnitude estimation function. \blue{Drawing lessons from the concept of Taylor's theorem in general function approximation}, we propose a novel all-neural SE system called \underline{\textbf{Ta}}ylor\underline{\textbf{E}}nhance\underline{\textbf{r}} (abbreviated as \textbf{TaEr}) to mimic the behavior of Taylor's series with respect to $X$ with the all-neural setting, and the recovery process can thus be modeled as the superimposition of the 0th-order and high-order polynomials, where the former only concerns the magnitude recovery and the latter polynomials are tasked with complex-residual estimation. \blue{Note that, unlike previous works where the entire network was encapsulated within a black box, this paper designs the modules of the network based on a specific mathematical tool. By conveniently increasing or decreasing the number of expansion orders, we can adjust the performance of the network depending on the parameter and computational complexity requirements. The proposed model is therefore highly flexible compared with existing models.}

\blue{In our preliminary conference paper~{\cite{li2022taylor}}, we  proposed an unfolding single-channel SE framework motivated by Taylor's theorem. However, the complex-valued characteristic of the spectrum was ignored in the derivation process.} In the journal version, we add several key modifications and contributions, which are summarized as four-fold:
\begin{enumerate}
	\item In the earlier version, the derivation is based on the scalar sense and the complex-valued property of the spectrum is neglected. Here we modify the defect and present a multi-variate mapping format of Taylor's series expansion.
	
	\item \blue{Despite the promising performance of advanced SE systems, they often suffer from high parameters and computational complexity, which can be impractical for real-world applications. To address this limitation, we propose \textbf{TaErLite}, a lightweight version of TaEr that achieves comparable performance to baselines while requiring substantially much fewer parameters, reducing computational complexity, and a shorter buffer length. Our approach offers potential for implementation in various real-time applications such as remote conferencing systems.}
	
	\item For practical devices, it is convenient and also cheap to place a microphone array. As such, we propose the multi-channel version of the proposed TaEr and TaErLite, which can utilize the spatial information to better suppress interferences.
	
	\item \blue{We present comprehensive ablation studies and quantitative comparisons with previous SOTA SE systems for both single-channel and multi-channel setups to demonstrate the superiority of the proposed method.}
\end{enumerate}
\section{Related Work}
\label{sec:related-works}
In this section, we first briefly review previous noise reduction methods in the T-F domain and time domain for the single-channel case. Then we briefly review existing multi-channel SE methods.
\subsection{Monaural Speech Enhancement Methods in the T-F Domain}
\label{sec:monaural-tf-domain-methods}
Before stepping into the deep-learning era, most SE algorithms were developed in the T-F domain since STFT provides a feasible analysis foundation. Traditional methods consist of spectral subtraction algorithm~{\cite{boll1979suppression}}, Wiener filtering~{\cite{sreenivas1996codebook}}, statistical signal processing-based methods~{\cite{scalart1996speech, ephraim1984speech}} and so on. Many of these methods only consider the spectral magnitude while neglecting the phase term as phase was ever thought to be unimportant~{\cite{wang1982unimportance}}. \blue{A recent work~{\cite{Landwehr}} also showed that phase information tends to be less useful in the speech separation task with heavy reverberation.} After that, Paliwal \emph{et al.}~{\cite{paliwal2011importance}} showed that the recovery of phase structure can improve the subjective perceptual quality. And many phase-aware methods~{\cite{krawczyk2014stft, gerkmann2015phase}} were subsequently proposed to incorporate the statistical distribution of the spectral phase, which achieve better performance over magnitude-only methods.

In the past several years, a multitude of complex-domain SE models were proposed. \blue{For instance, Williamson \emph{et al.}~{\cite{williamson2015complex}} proposed a complex-valued ratio mask (CRM) to filter the real and imaginary (RI) parts of the target spectrum, and Tan \emph{et al.}~{\cite{tan2019learning}} directly predicted the RI parts based on convolutional recurrent networks (CRNs), which can implicitly recover both the magnitude and phase information. In~{\cite{yin2020phasen}}, the magnitude and phase were tackled separately and a dual-path network is devised to estimate the magnitude and phase, respectively. Nonetheless, due to the wrapping effect of the phase, the phase branch tends to be sensitive toward the non-linear transform, which heavily limits the estimation capability.} \blue{Hu \emph{et al.} ~{\cite{hu2020dccrn}} proposed a complex-domain based convolutional-recurrent neural network, where both convolutional layers, normalization, activation, and long-short term memory (LSTM) layers simulate complex-valued operations.} In~{\cite{zheng2021interactive}}, Zheng \emph{et al.} proposed a two-branch network in the complex domain, where the information interaction between speech and noise branches is adopted to boost speech prediction. To alleviate the compensation effect between magnitude and phase, Li \emph{et al.}~{\cite{li2021two}} proposed a two-stage method to decompose the recovery problem into the magnitude and complex residual estimation. And a two-path estimation tactic was proposed~{\cite{li2022filtering}} to adopt the two steps in parallel. Based on attention-in-attention transformer (AIAT)~{\cite{vaswani2017attention}}, Yu \emph{et al.}~{\cite{yu2022dbt}} devised a federative network to estimate the magnitude and phase collaboratively with light-weight parameters. And Abdulatif \emph{et al.}~{\cite{abdulatif2022cmgan}} further improved the efficacy by replacing the AIAT with the conformer structure~{\cite{gulati2020conformer}}.
\subsection{Monaural Speech Enhancement Methods in the Time Domain}
\label{sec:monaural-time-domain-methods}
\vspace{-0.05cm}
Thanks to the powerful mapping capability of DNNs, a flurry of time domain-based methods were developed in recent years and gained domination in various front-end tasks. SEGAN~{\cite{pascual2017segan}} was the pioneering work to incorporate U-Net and generative adversarial network (GAN) to recover raw waveforms but the performance is just comparable to conventional Wiener filtering. In~{\cite{fu2018end}}, an utterance-based fully convolutional network (FCN) was proposed, which incorporated the metric and euclidean-distance loss. Pandey \emph{et al.}~{\cite{pandey2020densely}} proposed to stack multiple frames into a 2-D structure and learn it with a densely-connected 2-D U-Net~{\cite{ronneberger2015u}}. Except for directly learning the distribution of waveform samples, another typical strategy is to follow the ``Encoder-Separator-Decoder'' processing flow, where an over-complete feature encoding modules are utilized to convert the short-length waveform signals into the latent space in which multiple sources can be better separated. In~{\cite{luo2019conv}}, Conv-TasNet was proposed, where stacked temporal convolution networks (TCNs) serve as a separator. \blue{Later, a dual-path recurrent neural network (DPRNN) separator was proposed~{\cite{luo2020dual}}}, in which an intra- and inter-LSTM are to model the sequence relation in the short- and long-range, respectively. \blue{End-to-end time domain-based methods were criticized for their notably reduced performance in noisy-reverberant scenarios~{\cite{maciejewski2020whamr}}. However, a recent study by Heitkaemper \emph{et al.}~{\cite{Heitkaemper2020critic}} dissected the gains of TasNet by replacing internal components and indicated that the use of a learned latent space may not be the primary cause of performance degradation under reverberant conditions.}

\subsection{Multi-channel Speech Enhancement Methods}
\label{sec:multi-channel-methods}
Compared with the monaural setting, multi-channel SE algorithms can utilize the spatial cue to better distinguish different sound sources. A type of classical approach is called acoustic beamforming, where a set of linear filters are applied to the input mixtures to extract the signal of interest (SOI) and attenuate the interferences from other directions. Conventional beamforming methods include minimum variance distortionless response (MVDR) beamformer, multi-channel Wiener Filter (MWF) and so forth~{\cite{gannot2017consolidated}}. With the help of DNNs, NN-based multi-channel SE methods were developed in recent years~{\cite{heymann2016neural, luo2020end, wang2020complex, gu2021complex, ochiai2017unified, tan2022neural}}. A typical regime is to combine a general mask estimator with traditional beamformers. For example, Heymann \emph{et al.}~{\cite{heymann2016neural}} proposed to estimate the speech/noise ideal-binary mask (IBM) of each channel with bidirectional LSTMs, and the steering vector and noise spatial covariance matrix were derived, and the beamforming weights can then be calculated. Besides, some research works follow the ``extraction-fusion" paradigm, where the spectral and spatial features are first explicitly/implicitly extracted and sent to the fusion network. For example, Tan \emph{et al.}~{\cite{tan2022neural}} proposed to concatenate the spectra of different microphones across the channel dimension and send them to a densely-connected CRN for complex spectrum mapping, where the network serves as the spatial-spectral filter. Gu \emph{et al.}~{\cite{gu2021complex}} jointly optimized a complex-domain mask estimator and MVDR to reduce the nonlinear speech distortion introduced by the network. Besides, some efforts have been made to directly estimate the beamforming weights in either the time domain~{\cite{luo2020end, luo2022time}} or T-F domain~{\cite{xiao2016deep, 9596418, 9747528, li2022embedding, li2022taylorbeamformer, zhang2021multi}}, which strike a competitive trade-off between noise attenuation and speech distortion. \blue{In this paper, we extend the proposed model to the multi-channel case by jointly utilizing spatial and spectral features for better speech restoration.}
\section{Proposed Methodology}
\label{sec:problem-formulation}
In this section, we first demonstrate our method under the single-channel condition, then we extend it to the multi-channel setting.
\subsection{From Target Decoupling to Taylor's Series Expansion}
\label{sec:from-target-decoupling-to-taylor}
\vspace{-0.00cm}
When only a single microphone is available, the received noisy mixture in the T-F domain can be modeled as
\begin{equation}
\label{eqn1}
X_{l, k} = S_{l, k} + N_{l, k},
\end{equation}
where $X_{l, k} = \left|X_{l, k}\right|e^{j\theta_{X_{l, k}}}\in\mathbb{C}$, $S_{l, k} = \left|S_{l, k}\right|e^{j\theta_{S_{l, k}}}\in\mathbb{C}$, $N_{l, k}= \left|N_{l, k}\right|e^{j\theta_{N_{l, k}}}\in\mathbb{C}$ respectively denote the mixture, clean, and interference signals with the time index of $l\in\left\{1, \cdots, L\right\}$ and frequency index of $k\in\left\{1, \cdots, K\right\}$. $\left\{\theta_{X_{l, k}}, \theta_{S_{l, k}}, \theta_{N_{l, k}}\right\}$ denote the corresponding phases of the mixture, clean and interference, respectively. Note that when reverberation is not considered, \emph{i.e.}, in anechoic case, only environmental noise needs to be removed, and both noise and late reverberation components are regarded as the interference under the reverberant condition. Now onwards, if no confusion arises, we will omit the subscripts $\left\{k, l\right\}$. For SE task, the main goal is to recover the target speech from the input mixture, \emph{i.e.},
\begin{equation}
\label{eqn2}
\tilde{S} = \mathcal{F}\left(X\right),
\end{equation}
\blue{where $\mathcal{F}\left(\cdot\right)$ denotes the function to obtain the estimated clean speech and the tilde refers to the estimated variable.} For NN-based methods, the estimation process is usually encapsulated into a complicated network and thus has weak interpretability in the intermediate processing steps~{\cite{hu2020dccrn, tan2019learning}}. Recently, an intuitive decoupling-style estimation strategy was proposed~{\cite{li2021two, li2022filtering}}, in which two core operations are involved and summarized as
\begin{itemize}
	\item \textbf{Op1}: Only considering the magnitude term while ignoring the phase term to obtain a coarsely estimated complex spectrum.
	\item \textbf{Op2}: Facilitating the phase recovery by complex-valued residual estimation. 
\end{itemize}

Obviously, such a two-stage estimation pipeline follows the ``divide-and-conquer'' wisdom and provides us with a more flexible solution to tackle the problem progressively. In traditional SE algorithms, \textbf{Op1} can be fulfilled via either spectral substraction~{\cite{boll1979suppression}} or gain filtering such as Wiener filtering~{\cite{ephraim1984speech}}. \blue{And the above two-step scheme can be respectively written as}
\begin{equation}
\label{eqn3}
S = \left|\left|X\right|^{2} - \left|\tilde{N}\right|^{2}\right|^{0.5}e^{j\theta_{X}} + \underline{\delta}_{1},
\end{equation}
\begin{equation}
\label{eqn4}
S = \tilde{G}\left|X\right|e^{j\theta_{X}} + \underline{\delta}_{2},
\end{equation}
where $\tilde{\left|N\right|}$ and $\tilde{G}$ are the estimated noise spectral magnitude and real-valued gain function, respectively. \blue{For Eq.~(\ref{eqn3}), we usually assume the speech and noise components to be uncorrelated.} \blue{And $\left\{\underline{\delta}_{1}, \underline{\delta}_{2}\right\}\in\mathbb{C}$ are the corresponding complex-valued residual terms used to repair the phase.} Let $\tilde{\left|N\right|} = \sqrt{1 - \tilde{G}^{2}}\left|X\right|$, then Eqs.~(\ref{eqn3}) and (\ref{eqn4}) can have the similar format, as \textbf{Op1} is in essence to obtain an accurate noise power estimaiton~{\cite{gerkmann2011unbiased}}, which can then be removed from the mixture. Accordingly, Eqs.~(\ref{eqn3}) and (\ref{eqn4}) can be unified into
\begin{equation}
\label{eqn5}
S = \tilde{G}\left(\left|X\right|e^{j\theta_{X}} + \frac{\underline{\delta}}{\tilde{G}}\right) = \tilde{G}\left(X + \delta\right),
\end{equation}
\blue{where $\delta = \frac{\underline{\delta}}{\tilde{G}} = S\tilde{G}^{-1} - X$. Let us define $\hat{X} = S\tilde{G}^{-1} = \left|S\tilde{G}^{-1}\right|e^{j\theta_{S}}$, whch indicates the component before magnitude filtering. Here we analyze it with a special case, where $\tilde{G}$ is assumed to be an oracle magnitude filter, \emph{i.e.}, $\tilde{G} = \left|\frac{S}{X}\right|$. Then $\delta = S\tilde{G}^{-1} - X = \left|X\right|\left(e^{j\theta_{S}} - e^{j\theta_{X}}\right)$, which evaluates the phase gap between target speech and noisy mixture, is weighted by the mixture magnitude. \blue{Apparently, $\delta$ and $\tilde{G}$ are correlated and should be estimated simultaneously.} In traditional gain filtering methods, the estimation of the spectral gain usually involves the calculation of the prior SNR and noise power spectral density, which are based on noisy input spectrum. Therefore, the estimated filter $\tilde{G}$ should be related to the RI parts of the noisy spectrum, \emph{i.e.}, $\tilde{G}\left(X_{r}, X_{i}\right)$, and is assumed to meet the condition $\tilde{G}\left(X_{r}, X_{i}\right) = \tilde{G}\left(X_{i}, X_{r}\right)$, indicating that exchanging the order of $X_{r}$ and $X_{i}$ does not affect the value of the gain. Subscripts $\left\{r, i\right\}$ respectively denote the real and imaginary parts of the complex spectrum. This property is common and also reasonable in traditional gain-based methods as magnitude is usually calculated first and the phase component is dropped, \emph{i.e.}, $\left|X\right| = \sqrt{X_{r}^{2} + X_{i}^{2}}$.}

\blue{Further, let us abstract the filtering operation by $\tilde{G}$ into a general magnitude filtering function $\mathcal{G}$. Eq.~({\ref{eqn5}}) can thus be split into the RI parts~{\cite{williamson2015complex}} and written as} 
\begin{equation}
\label{eqn6}
\blue{S_{r} = \mathcal{G}\left(X_{r} + \delta_{r}, X_{r} + \delta_{r_{1}}, X_{i} + \delta_{i_{2}}\right)},
\end{equation}
\begin{equation}
\label{eqn7}
\blue{S_{i} = \mathcal{G}\left(X_{i} + \delta_{i}, X_{i} + \delta_{i_{3}}, X_{r} + \delta_{r_{4}}\right)},
\end{equation}
\blue{where $\mathcal{G}$ is defined as a trivariate function and can be expressed as}
\begin{equation}
\label{eqn8}
\blue{\mathcal{G}\left(a, b, c\right) = a\tilde{G}\left(b, c\right).}
\end{equation}

\blue{Comparison of Eqs.~({\ref{eqn6}}) - ({\ref{eqn7}}) and Eq.~({\ref{eqn8}}) shows that Eqs.~({\ref{eqn6}}) - ({\ref{eqn7}}) are special cases of Eq.~({\ref{eqn8}}) when $\left\{a = X_{r} + \delta_{r}, b = X_{r} + \delta_{r_{1}}, c = X_{i} + \delta_{i_{2}}\right\}$ and $\left\{a = X_{i} + \delta_{i}, b = X_{i} + \delta_{i_{3}}, c = X_{r} + \delta_{r_{4}} \right\}$, respectively, and $\left\{\delta_{r}, \delta_{r_{1}}, \delta_{i_{2}}, \delta_{i}, \delta_{i_{3}}, \delta_{r_{4}}\right\}$ are the corresponding unknown sparse terms. Since the estimation of the magnitude filter involves the RI of the input mixture, $\left\{\delta_{r_{1}}, \delta_{i_{2}}, \delta_{i_{3}}, \delta_{r_{4}} \right\} = 0$, which means that only the neighborhood point of the variable $a$ in $\mathcal{G}\left(a, b, c\right)$ is involved and can be regarded as the special case of a general trivariate function. Therefore, we can rewrite Eqs.~({\ref{eqn6}}) - ({\ref{eqn7}}) as}
\begin{equation}
\label{eqn9}
\blue{S_{r} = \mathcal{G}\left(X_{r}+\delta_{r}, X_{r}, X_{i}\right) = \tilde{G}\left(X_{r}, X_{i}\right)\left[X_{r} + \delta_{r}\right]}
\end{equation}
\begin{equation}
\label{eqn10}
\blue{S_{i} = \mathcal{G}\left(X_{i}+\delta_{i}, X_{i}, X_{r}\right) = \tilde{G}\left(X_{i}, X_{r}\right)\left[X_{i} + \delta_{i}\right]}
\end{equation}

\blue{As $\left\{\delta_{r}, \delta_{i}\right\}$ are unknown in practice, we actually cannot resolve the mapping function of $\left\{X_{r} + \delta_{r}, X_{r}, X_{i}\right\}$ in Eq.~({\ref{eqn9}}) and that of $\left\{X_{i} + \delta_{i}, X_{i}, X_{r}\right\}$ in Eq.~({\ref{eqn10}}) directly. Toward this end, we resort to Taylor's theorem, which is widely utilized for general function approximation. Specifically, we assume the function $\mathcal{G}\left(X_{r} + \delta_{r}, X_{r}, X_{i}\right)$ and $\mathcal{G}\left(X_{i} + \delta_{i}, X_{i}, X_{r}\right)$ to be differentiable with respect to (w.r.t.) $Q$ orders, Eqs.~({\ref{eqn9}}) - ({\ref{eqn10}}) can then be further written as}
\begin{equation}
\begin{aligned}[b]
\label{eqn12}
\blue{S_{r}} &\blue{\approx \mathcal{G}\left(X_{r}, X_{r}, X_{i}\right) + \frac{1}{1!}\frac{\partial\mathcal{G}\left(X_{r}, X_{r}, X_{i}\right)}{\partial X_{r}}\delta_{r} + \cdots} \\
   & \blue{+ \frac{1}{Q!}\frac{\partial^{Q}\mathcal{G}\left(X_{r}, X_{r}, X_{i}\right)}{\partial^{Q} X_{r} }\delta_{r}^{Q}},
\end{aligned}
\end{equation}
\begin{equation}
\begin{aligned}[b]
\label{eqn13}
\blue{S_{i}} &\blue{\approx \mathcal{G}\left(X_{i}, X_{i}, X_{r}\right) + \frac{1}{1!}\frac{\partial\mathcal{G}\left(X_{i}, X_{i}, X_{r}\right)}{\partial X_{i}}\delta_{i} + \cdots} \\
& \blue{+ \frac{1}{Q!}\frac{\partial^{Q}\mathcal{G}\left(X_{i}, X_{i}, X_{r}\right)}{\partial^{Q} X_{i} }\delta_{i}^{Q}},
\end{aligned}
\end{equation}
\blue{where $\mathcal{G}\left(X_{r}, X_{r}, X_{i}\right) = X_{r}\tilde{G}\left(X_{r}, X_{i}\right)$, $\mathcal{G}\left(X_{i}, X_{i}, X_{r}\right) = X_{i}\tilde{G}\left(X_{i}, X_{r}\right)$, and Eqs.~({\ref{eqn12}})~-~({\ref{eqn13}}) can be simplified to}
\begin{equation}
\label{eqn14}
\blue{S_{r} \approx \underbrace{\mathcal{G}\left(X_{r}, X_{r}, X_{i}\right)}_{\text{0th-order term}} + \underbrace{\sum_{q=1}^{Q}\frac{1}{q!}\mathcal{H}_{r}\left(q, X, \delta_{r} \right)}_{\text{high-order terms}}},
\end{equation}
\begin{equation}
\label{eqn15}
\blue{S_{i} \approx \underbrace{\mathcal{G}\left(X_{i}, X_{i},  X_{r}\right)}_{\text{0th-order term}} + \underbrace{\sum_{q=1}^{Q}\frac{1}{q!}\mathcal{H}_{i}\left(q, X, \delta_{i} \right)}_{\text{high-order terms}}},
\end{equation}
\blue{where}
\begin{equation}
\label{eqn16}
\blue{\mathcal{H}_{r}\left(q, X, \delta_{r}\right) = \frac{\partial^{q}\mathcal{G}\left(X_{r}, X_{r}, X_{i}\right)}{\partial^{q} X_{r}}\delta^{q}_{r}},
\end{equation}
\begin{equation}
\label{eqn17}
\blue{\mathcal{H}_{i}\left(q, X, \delta_{i}\right) = \frac{\partial^{q}\mathcal{G}\left(X_{i}, X_{i}, X_{r}\right)}{\partial^{q} X_{i}}\delta^{q}_{i}},
\end{equation}

\blue{Note that when $Q$ is set to 0, $\mathcal{H}_{r}\left(0, X, \delta_{r}\right) = \mathcal{G}\left(X_{r}, X_{r}, X_{i}\right)$ and $\mathcal{H}_{i}\left(0, X, \delta_{i}\right) = \mathcal{G}\left(X_{i}, X_{i}, X_{r}\right)$. As such, we provide a novel perspective to demonstrate the magnitude-phase decoupling strategy of the complex spectrum recovery process. Concretely, in the 0th-order term, only the magnitude function is considered, and therefore, it serves as a denoiser to suppress the interfernce in the magnitude domain. For high-order terms, as $\left\{\delta_{r}, \delta_{i}\right\}$ are involved in the calculation, they operate in the complex domain and are tasked for complex spectrum refinement. From Eqs.~(\ref{eqn12}) and (\ref{eqn13}), one can see that they have a similar format. Without loss of generality, in the remaining derivation, we take the recovery of the real component of the target complex spectrum as an example and a similar process can be applied to the imaginary part.}
\subsection{Relations between Adjacent High-Order Terms}
\label{sec:correlations-between-adjacent-order-terms}
\blue{To enable the estimation of high-order terms, it is imperative to investigate the relation between adjacent order terms. Let us differentiate $\mathcal{H}_{r}\left(q, X, \delta_{r}\right)$ w.r.t. $X_{r}$, given by}
	\begin{equation}
	\label{eqn18}
	\begin{aligned}[b]
	\blue{\frac{\partial\mathcal{H}_{r}\left(q, X, \delta_{r}\right)}{\partial X_{r}}} &\blue{= \frac{\partial}{\partial X_{r}}\left(\frac{\partial^{q}\mathcal{G}\left(X_{r}, X_{r}, X_{i}\right)}{\partial^{q} X_{r}}\delta^{q}_{r}\right)}\\
	& \blue{= \frac{\partial^{\left(q+1\right)}\mathcal{G}\left(X_{r}, X_{r},  X_{i}\right)}{\partial^{\left(q+1\right)}}\delta^{q}_{r}}\\
	& \blue{+ q\frac{\partial^{q}\mathcal{G}\left(X_{r}, X_{r}, X_{i}\right)}{\partial^{q}X_{r}}\delta_{r}^{\left(q-1\right)}\left(\frac{\partial\delta_{r}}{\partial X_{r}}\right)}.
	\end{aligned}
	\end{equation}
	
\blue{Recall that}
\begin{gather}
\label{eqn19}
\blue{\frac{\partial \delta_{r}}{\partial X_{r}} = \frac{\partial\left(\hat{X}_{r} - X_{r}\right)}{\partial X_{r}} = \frac{\partial \hat{X}_{r}}{\partial X_{r}} - 1}.
\end{gather}

\blue{Let us denote $\mathcal{K}\left(\hat{X}_{r}, X_{r}\right) = \frac{\partial \hat{X}_{r}}{\partial X_{r}}$. Multiplying $\delta_{r}$ on both side of Eq.~({\ref{eqn18}}) and substituting Eq.~({\ref{eqn19}}) into Eq.~({\ref{eqn18}}), we obtain}
\begin{equation}
\label{eqn20}
\begin{aligned}[b]
\blue{\mathcal{H}_{r}\left(q+1, X, \delta_{r}\right)} &\blue{= q\mathcal{H}_{r}\left(q, X, \delta_{r}\right) + \frac{\partial\mathcal{H}_{r}\left(q, X, \delta_{r}\right)}{\partial X_{r}}\delta_{r}}\\
& \blue{- q\mathcal{H}_{r}\left(q, X, \delta_{r}\right)\mathcal{K}\left(\hat{X}_{r}, X_{r}\right)}.
\end{aligned}
\end{equation}

\blue{For $\mathcal{H}_{i}\left(q, X, \delta_{i}\right)$, a similar method can be utilized and we can derive the recursive relation between $H_{i}\left(q+1, X, \delta_{i}\right)$ and $H_{i}\left(q, X, \delta_{i}\right)$, given by}
\begin{equation}
\label{eqn21}
\begin{aligned}[b]
	\blue{\mathcal{H}_{i}\left(q+1, X, \delta_{i}\right)} &\blue{= q\mathcal{H}_{i}\left(q, X, \delta_{i}\right) + \frac{\partial\mathcal{H}_{i}\left(q, X, \delta_{i}\right)}{\partial X_{i}}\delta_{i}}\\
	& \blue{- q\mathcal{H}_{i}\left(q, X, \delta_{i}\right)\mathcal{K}\left(\hat{X}_{i}, X_{i}\right)}.
\end{aligned}
\end{equation}
\subsection{Taylor's Series Expansion in the Multi-channel Case}
\label{sec:extension-to-multi-channel-case}
\blue{In the above part, we derive the recursive relation between neighboring high-order polynomials in the single-channel case. In practical applications, it is common and simple to deploy a microphone array on the speech interaction devices, which can provide effective discrimination among different acoustic sources with the spatial information. Therefore, we extend the Taylor modeling into the multi-channel case for more general purposes. Under multi-channel condition, the recording signals have $M$ channels, which can be formulated as $\mathbf{X}_{l, k} = \mathbf{S}_{l, k} + \mathbf{N}_{l, k}$, where $\left\{\mathbf{X}_{l, k}, \mathbf{S}_{l, k}, \mathbf{N}_{l, k}\right\}\in\mathbb{C}^{M}$ and $M$ denotes the number of microphones. Without losing generality, the first channel is chosen as the reference channel. \blue{Similar to Eqs.~(9) and (10), we regard the complex spectra from different channels as the inputs of the magnitude function so that the general function can fuse both spectral and spatial cues to recover the target speech of the reference microphone~{\cite{tan2022neural}}.} Then the estimation of $\left\{S_{r}, S_{i}\right\}$ can be expressed as}
\begin{equation}
\label{eqn22}
\begin{aligned}[b]
\blue{S_{r}} &\blue{= \mathcal{G}\left(X_{r}^{(1)}+\delta_{r}, X_{r}^{(1)}, X_{i}^{(1)},\cdots,X_{r}^{(M)},X_{i}^{(M)}\right)}\\
&\blue{= \tilde{G}\left(X_{r}^{(1)},X_{i}^{(1)},\cdots,X_{r}^{(M)},X_{i}^{(M)}\right)\left[X_{r}^{\left(1\right)} + \delta_{r}\right]},
\end{aligned}
\end{equation}
\begin{equation}
\label{eqn23}
\begin{split}
\blue{S_{i}} &\blue{= \mathcal{G}\left(X_{i}^{(1)}+\delta_{i}, X_{i}^{(1)}, X_{r}^{(1)},\cdots,X_{i}^{(M)},X_{r}^{(M)}\right)}\\
&\blue{= \tilde{G}\left(X_{i}^{(1)},X_{r}^{(1)},\cdots,X_{i}^{(M)},X_{r}^{(M)}\right)\left[X_{i}^{\left(1\right)} + \delta_{i}\right]},
\end{split}
\end{equation}
\blue{where $\mathcal{G}$ is defined as a magnitude filtering function with $2M + 1$ input variables. $\left(\cdot\right)^{\left(m\right)}$ denotes the channel index. With Taylor's theorem, one can get}
\begin{equation}
\label{eqn24}
\blue{S_{r} \approx \mathcal{G}\left(\mathbf{X}_{r}, \mathbf{X}_{i}\right)
+ \sum_{q=1}^{Q}\frac{1}{q!}\mathcal{H}_{r}\left(q, \mathbf{X}, \delta_{r} \right)},
\end{equation}
\begin{equation}
\label{eqn25}
\blue{S_{i} \approx \mathcal{G}\left(\mathbf{X}_{i}, \mathbf{X}_{i}\right)
+ \sum_{q=1}^{Q}\frac{1}{q!}\mathcal{H}_{i}\left(q, \mathbf{X}, \delta_{i} \right)},
\end{equation}
\blue{where} 
\begin{equation}
\label{eqn26}
\blue{\mathcal{H}_{r}\left(q, \mathbf{X}, \delta_{r} \right) = \frac{\partial^{q}\mathcal{G}\left(\mathbf{X}_{r}, \mathbf{X}_{i}\right)}{\partial^{q} X^{\left(1\right)}_{r}}\delta^{q}_{r}},
\end{equation}
\begin{equation}
	\label{eqn27}
	\blue{\mathcal{H}_{i}\left(q, \mathbf{X}, \delta_{i} \right) = \frac{\partial^{q}\mathcal{G}\left(\mathbf{X}_{i}, \mathbf{X}_{r}\right)}{\partial^{q} X^{\left(1\right)}_{i}}\delta^{q}_{i}}.
	\end{equation}

\blue{Similar to Eqs.~({\ref{eqn18}})~-~({\ref{eqn20}}), one can derive the relation between adjancent high-order terms as}
\begin{equation}
	\label{eqn28}
	\begin{aligned}[b]
	\blue{\mathcal{H}_{r}\left(q+1, \mathbf{X}, \delta_{r}\right)} &\blue{= q\mathcal{H}_{r}\left(q, \mathbf{X}, \delta_{r}\right) + \frac{\partial\mathcal{H}_{r}\left(q, \mathbf{X}, \delta_{r}\right)}{\partial X_{r}}\delta_{r}}\\
	& \blue{- q\mathcal{H}_{r}\left(q, \mathbf{X}, \delta_{r}\right)\mathcal{K}\left(\hat{X}_{r}^{\left(1\right)}, X_{r}^{\left(1\right)}\right)},
	\end{aligned}
\end{equation}

\begin{equation}
	\label{eqn29}
	\begin{aligned}[b]
	\blue{\mathcal{H}_{i}\left(q+1, \mathbf{X}, \delta_{i}\right)} &\blue{= q\mathcal{H}_{i}\left(q, \mathbf{X}, \delta_{i}\right) + \frac{\partial\mathcal{H}_{i}\left(q, \mathbf{X}, \delta_{i}\right)}{\partial X_{i}}\delta_{i}}\\
	& \blue{- q\mathcal{H}_{i}\left(q, \mathbf{X}, \delta_{i}\right)\mathcal{K}\left(\hat{X}_{i}^{\left(1\right)}, X_{i}^{\left(1\right)}\right)}.
	\end{aligned}
\end{equation}

From Eqs.~(\ref{eqn20}) and (\ref{eqn21}) and Eqs.~(\ref{eqn28}) and (\ref{eqn29}), one can get that there exist \blue{two complicated terms} with derivative operation on the right hand of the equations, which often causes calculation instability. Besides, \blue{as $\delta$ is assumed to exist and is introduced in advance, its actual distribution tends to vary in different scenarios}. To this end, we propose to replace the complicated terms using a surrogate function with trainable parameters, where the distribution can also be implicitly learned from the training data. Furthermore, we propose to train the whole system in an end-to-end manner, where both the general magnitude filtering function $\mathcal{G}$ and surrogate functions are jointly learned. Thus the system can mitigate the latent mismatch between adjacent order terms.
\begin{figure}[t]
	\renewcommand\arraystretch{0.5}
	\begin{algorithm}[H]
		\caption{Forward Processing Stream of the Proposed Taylor-Unfolding Framework for the Single-Channel Case.}
		\label{alg:mcd}
		\begin{algorithmic}[1]
			\STATE \textbf{Input:}\\
			\STATE \hspace{4mm} Noisy mixture $X = \left\{X_{r}, X_{i}\right\}$;\\
			\STATE \hspace{4mm} The number of the recursive order $Q$;\\
			\STATE \hspace{4mm} Current target estimation $Esti \gets 0$;\\
			\STATE \textbf{Output:}\\
			\STATE \hspace{4mm} The final estimated target $\tilde{S} = \left\{S_{r}, S_{i}\right\}$;
			\STATE $\mathbf{R} = High\text{-}Order Encoder\left(X_{r}, X_{i}\right)$\\
			\STATE \textbf{for} $q = 0, ..., Q$ \textbf{do}\\
			\STATE \hspace{4mm} \textbf{if} $q = 0$ \textbf{then} \\
			\STATE \hspace{8mm} $\mathcal{H}_{r}\left(0, X, \delta_{r}\right) = \mathcal{G}\left(X_{r}, X_{r}, X_{i}\right)$\\
			\STATE \hspace{8mm} $\mathcal{H}_{i}\left(0, X, \delta_{i}\right) = \mathcal{G}\left(X_{i}, X_{i}, X_{r}\right)$\\
			\STATE \hspace{8mm} $Esti \gets Esti + \frac{1}{0!}\mathcal{H}\left(0, X, \delta\right)$
			\STATE \hspace{8mm} \blue{$\mathcal{H}\left(q, X, \delta\right) = \left(q - 1\right)\mathcal{H}\left(q - 1, X, \delta\right) + \mathcal{P}\left(q-1, \mathcal{H}, \mathbf{R}\right)$}\\
			\STATE \hspace{8mm} \normalsize $Esti \gets Esti + \frac{1}{q!}\mathcal{H}\left(q, X, \delta\right)$\\
			\STATE \hspace{4mm} \textbf{end if}\\
			\STATE \textbf{end for}\\
			\STATE $\tilde{S} \gets Esti$ 
		\end{algorithmic}
	\end{algorithm}
	\vspace{-.2cm}
\end{figure}
\begin{figure*}[t]
	\centering
	\centerline{\includegraphics[width=1.8\columnwidth]{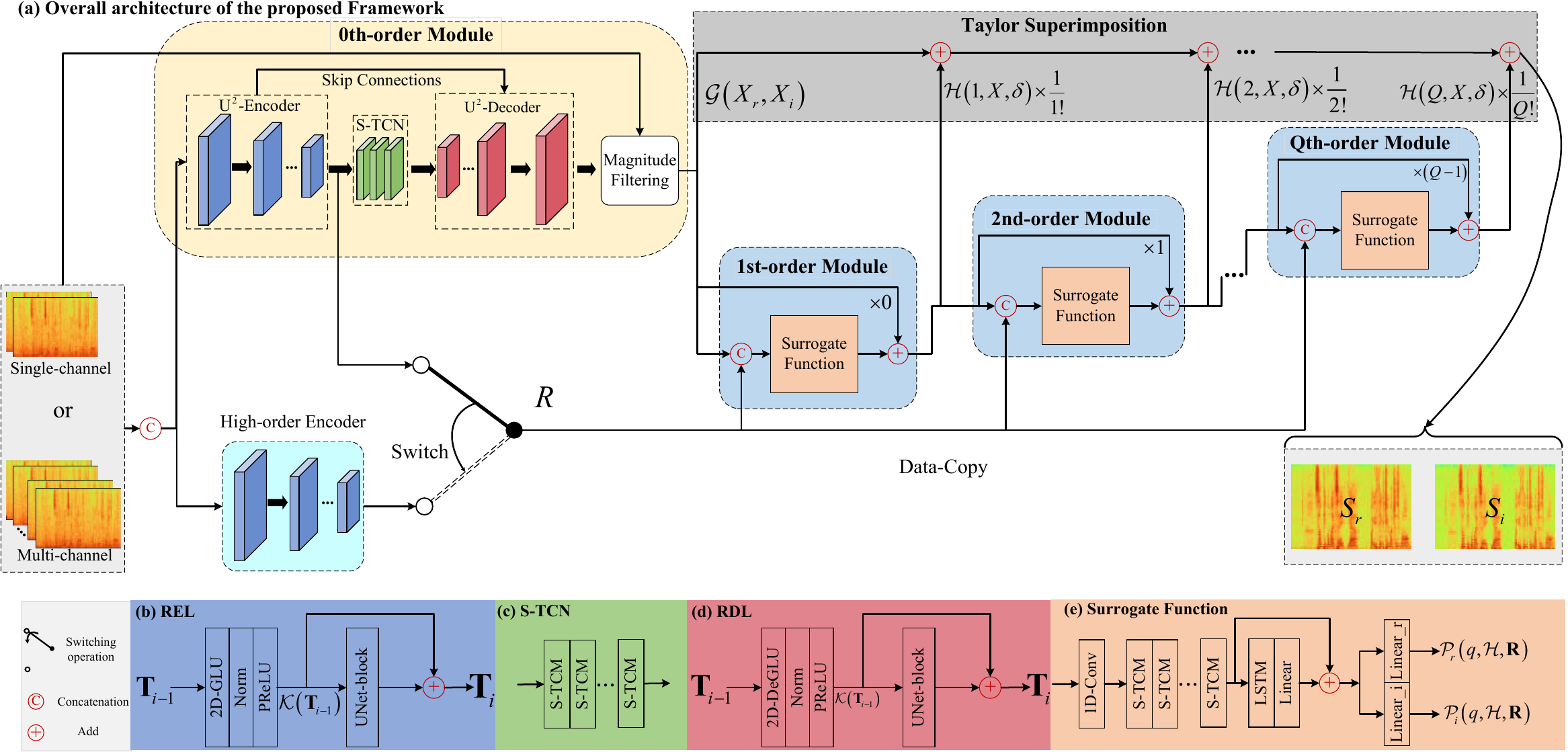}}
	\caption{(a) Overall network architecture of the proposed approach. Similar to Taylor's series expansion, it consists of the 0th-order term and multiple high-order terms, where the 0th-order term aims to coarsely remove the noise in the magnitude domain and high-order terms estimate the complex-valued residual components and are responsible for spectrum refinement. Following the calculation rule of Taylor superimposition, the spectrum of the target speech can be recovered. \blue{The input can be either RI for the single-channel case or the concatenation of RIs from different channels for the multi-channel case.} (b) Detailed structure of the REL. (c) Detailed structure of the S-TCN. (d) Detailed structure of the RDL. (e) Internal structure of the surrogate function, which \blue{replaces the original complicated terms}. Different modules are indicated with different colors.}
	\label{fig:architecture}
	\vspace{-0.3cm}
\end{figure*}
\subsection{Proposed Taylor-Unfolding Framework}
\label{sec:proposed-taylor-unfolding-framework}
The overall diagram of the proposed TaEr is shown in Fig.~{\ref{fig:architecture}}(a), which simulates the calculation flow of Taylor's series expansion. According to our previous formulation, \blue{for the single-channel case}, the RI parts are viewed as the input of the general function, so we concatenate RI along the channel dimension as the network input. And if a microphone array is deployed, the RI parts of different microphones will be stacked. As a result, the model can seamlessly support both single-channel and multi-channel scenarios. The 0th-order module is devised as a neural magnitude filter to suppress the interference in the magnitude domain. As the calculation of the high-order terms involves the noisy input, a high-order encoder can be utilized to extract the features $\mathbf{R}$ from the input and is concatenated with the output of the last high-order module as the input of the current module. It is worth emphasizing that in the above part, the derivations of $\left\{\mathcal{H}_{r}, \mathcal{H}_{i}\right\}$ are independent, which means that ideally, we need two sets of trainable parameters to reconstruct the RI spectra, respectively. Nevertheless, benefiting from the multi-task learning characteristic of the neural network, \blue{we merge the 0th-order filtering module and the surrogate functions w.r.t. $\left\{\mathcal{H}_{r}, \mathcal{H}_{i}\right\}$ into a set of network, where the latter is denoted as  $\mathcal{P}\left(q, \mathcal{H}, \mathbf{R}\right)$}, \blue{and two separate output linear layers are exploited to generate the corresponding real and imaginary parts of the surrogate functions}, as shown in Fig.~{\ref{fig:architecture}}(e). After all the terms are obtained recursively, following the rule of Taylor's series expansion, we superimpose them together and output the final estimation. Taking the single-channel case as an example, the forward stream is summarized in Algorithm~{\ref{alg:mcd}}.

In the 0th-order module, we adopt a classical U-Net structure~{\cite{ronneberger2015u}}, where the encoder gradually decreases the feature size by consecutive downsampling operations, and the decoder can gradually recover the original feature size via multiple transposed convolution layers. Note that the time size remains unchanged during the process to guarantee the frame-level processing mechanism. However, excessive feature downsampling operations may lead to acoustic information blurring and loss, which can severely hinder the quality of the recovered speech spectrum. Motivated by the recent success of U$^{2}$-Net in the salient object detection field~{\cite{qin2020u2}}, we adopt a composite version of U-Net, where the encoder and decoder are comprised of multiple recalibration encoding layers (REL) and recalibration decoding layers (RDL), as shown in panels (b) and (d) of Fig.~{\ref{fig:architecture}}. Taking the $i$th REL as an example, the input feature first passes a 2D-gated linear unit (GLU)~{\cite{tan2019learning}}, cumulative layer normalization (cLN), and parametric ReLU (PReLU), then a UNet-block is exploited with residual learning, \blue{whose structure is similar to U-Net but the layer depth can dynamically change depending on the current feature size}. For example, \blue{if the depth of the $i$th UNet-block is denoted as $U_{i}$}, then $U_{i} = 3$ means that the UNet-block includes three encoding and decoding layers, respectively. RDL is similar to REL except the 2D-GLU is replaced by the transposed counterpart. To facilitate the long-term sequence modeling, squeezed TCN (S-TCN)~{\cite{li2021two}} is employed in the bottleneck, which includes multiple squeezed temporal convolution modules (S-TCMs) and can utilize the correlations between adjacent frames via dilated convolutions. The sigmoid function is used to restrict the gain value into $\left(0, 1\right)$.

For high-order terms, a high-order encoder is adopted for feature extraction from the input. To reduce the parameter redundancy, we attempt to directly utilize the encoded features from the 0th-order module to replace that of the high-order encoder, and no performance drop is observed, which will be shown in Section~{\ref{sec:ablation-study}}. The internal structure of the surrogate function is shown in Fig.~{\ref{fig:architecture}}(e). It includes a 1D-Conv, followed by stacked S-TCMs and a residual LSTM~{\cite{nicolson2019deep}}. Two linear layers are utilized to estimate the RI components. Notice that in this study, we do not focus on delicate structure design, and all the utilized network modules are typical and ever exist in the previous SE works. \blue{We believe that there exist more advanced modules to promote the overall performance, which is out of the scope  of this paper.}
\subsection{\blue{Relations between the Proposed Framework and Previous Work}}
\label{sec:relation}
\blue{We have to point out that we do not intend to pursue the rigorous simulation of Taylor's expansion since the end-to-end training strategy makes the intermediate estimation relatively uncontrollable and untractable. Nonetheless, this work proposes a different perspective on SE framework design. The proposed framework is essentially a multi-stage paradigm, where the 0th-order module recovers the speech component coarsely in the magnitude domain, and the higher-order modules compensate for residual estimation errors existed in earlier stages. These high-order modules are recursively updated using Eqs.~({\ref{eqn20}})~-~({\ref{eqn21}}) and Eqs.~({\ref{eqn28}})~-~({\ref{eqn29}}), and this approach is thus closely related to the recently proposed deep unfolding framework{\cite{hershey,wisdom}}, in which the performance can be improved by unfolding iterations as layers in a deep network. By setting the expansion order to 1, \emph{i.e.}, $Q = 1$, the proposed framework only covers two terms: the 0th- and the 1st-order modules, similar to the filtering-refining (FR) strategy described in~{\cite{li2022filtering}}. Thus, the proposed method can be regarded as a generalized case of the FR tactic. Additionally, as the proposed framework decouples the estimation of the magnitude and phase based on Taylor's series expansion, the former can be realized with a more light-weight method, \emph{e.g.}, Mel or equivalent rectangular bandwidth (ERB) domain. As such, it can be effectively employed in the resource-limited devices{\cite{deepfilter}}, which will be illustrated in Section~{\ref{sec:delve-into-light-weight-design-of-the-proposed-framework}}. In contrast, for existing complex spectral mapping methods, where the real and imaginary spectrograms are jointly estimated, the magnitude and phase components are coupled and need to be tackled simultaneously in the linear frequency scale, which often lead to relatively large computational complexity and may hinder flexibility in practical deployment.}
\subsection{Toward Light-weight Design of the Proposed Model}
\label{sec:delve-into-light-weight-design-of-the-proposed-framework}
\vspace{-0.05cm}
\blue{Admittedly, existing top-performing SE models have achieved gratifying performance in recent years, a majority of them, however, still face intrinsic challenges to deploy in the resource-restricted and real-time devices, which can be roughly attributed to three-fold}:
\begin{itemize}
	\item Large trainable parameters: Most T-F domain-based models operate on the fine-grained spectrum with a relatively large window length (\emph{e.g.}, 20-32 ms), so large trainable parameters are usually needed to tackle each T-F bin.
	\item Large computing complexity: the U-Net structure with skip-connection is widely adopted in the existing SE models~{\cite{tan2019learning, hu2020dccrn, li2021two}}, which is resource-consuming.
	\item Large buffer length: \blue{Some advanced modules like TCNs~{\cite{bai2018empirical}}, transformers~{\cite{vaswani2017attention}} and their variants can effectively establish long/global-term sequence relations, but a large history buffer is often needed}, which is hardware-unfriendly and can dominate the memory usage.
\end{itemize}

\begin{figure}[t]
	\centering
	\centerline{\includegraphics[width=0.88\columnwidth]{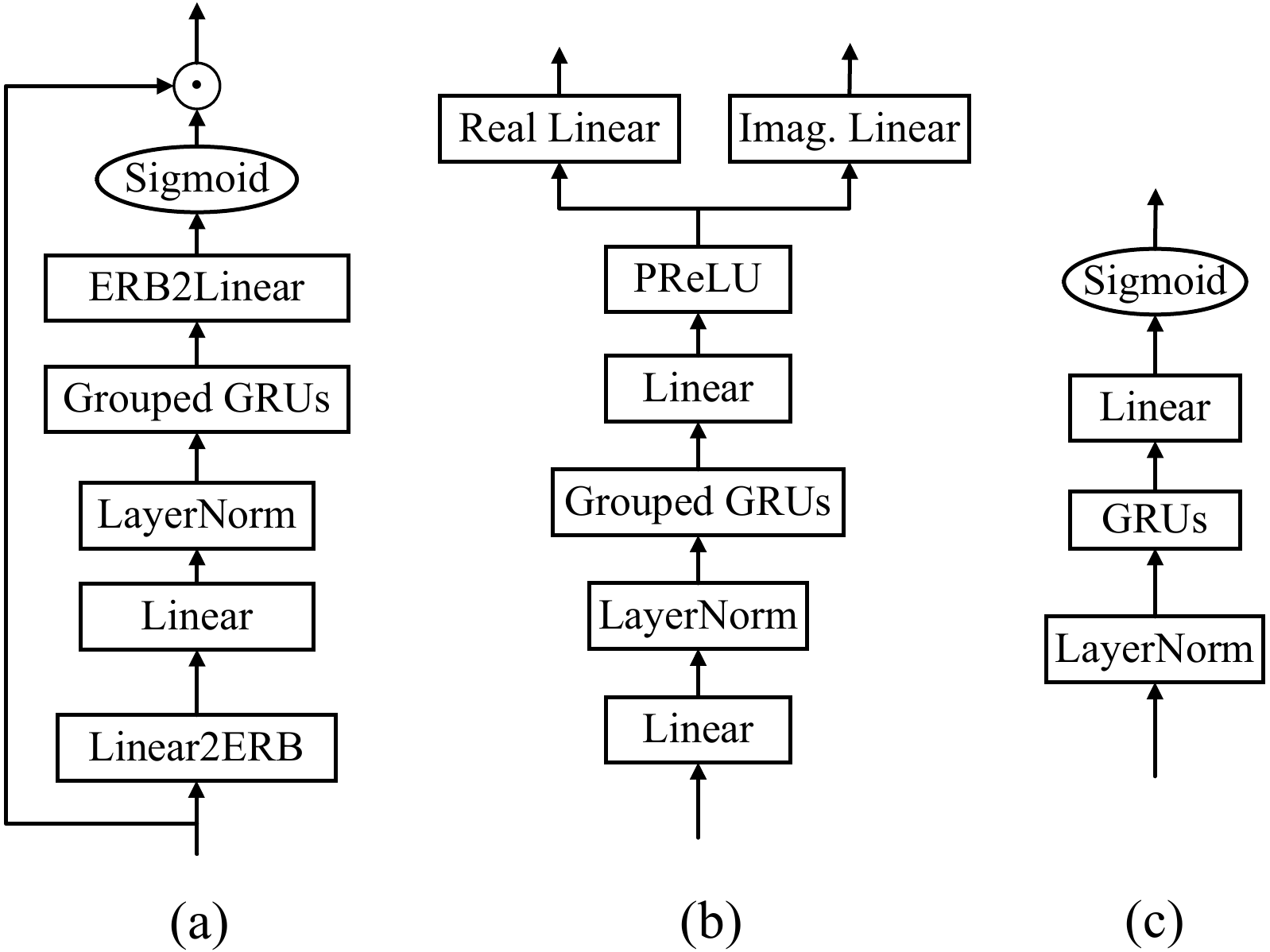}}
	\caption{\blue{(a) Internal structure of the 0th-order module in TaErLite}, ``Linear2ERB'' and ``ERB2Linear" respectively denote the linear transform from STFT domain to ERB domain and the linear transform from ERB domain to STFT domain. (b) Internal structure of the high-oder module in TaErLite. (c) Internal structure of the auxiliary post-filter.}
	\label{fig:taerlite-module}
	\vspace{-0.3cm}
\end{figure}

To mitigate the abovementioned problems, we propose TaErLite, \blue{a light-weight counterpart of TaEr} by replacing the corresponding parameter-, computation-, and buffer-costing modules with more light-weight operators, whose modifications are summarized as
\begin{itemize}
	\item In the 0th-order module, the frequency dimension of the input is largely reduced via the ERB filter-bank, in which the neighboring frequency bins are merged with the perceptual criterion~{\cite{deepfilter}}. Besides, we replace the original U-Net structure with two light-weight grouped GRU layers to reduce the parameters and computational complexity. Detailed structure is shown in Fig.~{\ref{fig:taerlite-module}}(a).
	\item To facilitate the phase recovery, a separate high-order encoder operated in the linear-scale frequency band is utilized but we drop all the UNet-blocks and also halve the channel number of intermediate 2D-Convs.
	\item In the high-order module, we adopt two grouped GRU layers with two output linear layers to model the surrogate function. In this way, only one past frame is required, and thus the large-buffer dependency of the staked dilated convolutions can be alleviated. Detailed structure is presented in Fig.~{\ref{fig:taerlite-module}}(b).
	\item To effectively suppress the residual noise caused by the coarse frequency modeling, an auxiliary post-filter is adopted after the superimposition, in which two tiny GRU layers are used to estimate the gain value at the frame level rather than at the bin level. Detailed structure is presented in Fig.~{\ref{fig:taerlite-module}}(c). \blue{Comparisons in terms of trainable parameters, computational complexity and time receptive field size between TaEr and TaErLite will be analyzed in Section~{\ref{tbl:taer-taerlite}}.}  
\end{itemize}
\begin{table*}[!t]
	\renewcommand\arraystretch{0.72}
	\small
	\centering
	\caption{Objective results of the ablation studies in terms of PESQ, eSTOI, and SI-SNR on the test set of WSJ0-SI84. ``$Q$'' denotes the number of expansion orders. ``High-order Encoder Shared'' denotes whether the extracted feature $\mathbf{R}$ comes from the reused encoded feature of the 0th-order module or not. ``Encoder Type'' denotes the network type in the 0th-order module and high-order encoder, and ``U$^{2}$'' and ``U'' namely denote U$^{2}$-Net and UNet. ``Acti. Func.'' denotes the activation function type, and ``P'', ``E'' refer to PReLU and ELU, respectively. The best and second best results are \textbf{boldfaced} and \underline{underlined}, respectively.}
	\label{tbl:ablation-study}
	\resizebox{0.92\textwidth}{!}{
		\begin{tabular}{c|cccccccccc}
			\toprule
			\multirow{2}*{IDs} &\multirow{2}*{System} &\multirow{2}*{$Q$} &High-order &Encoder &Acti. &Param. &MACs &\multirow{2}*{PESQ $\uparrow$} &\multirow{2}*{eSTOI (\%) $\uparrow$} &\multirow{2}*{SI-SNR (dB) $\uparrow$} \\
			& & &Encoder Shared &Type &Func. &(M) &(G/s) & & & \\
			\hline
			1 &TaEr &0  &\Checkmark  &U$^{2}$ &P &2.17 &3.93 &2.65 &70.60 &8.93\\
			2 &TaEr &1 &\Checkmark  &U$^{2}$ &P &3.59 &4.07 &2.77 &75.27 &10.63\\
			3 &TaEr &2 &\Checkmark  &U$^{2}$ &P &5.00 &4.22 &2.80 &76.19 &10.92\\
			4 &TaEr &3 &\Checkmark &U$^{2}$ &P &6.42 &4.36 &\underline{2.81} &76.32 &10.97\\
			5 &TaEr &4 &\Checkmark  &U$^{2}$ &P &7.83 &4.50 &\textbf{2.83} &\underline{76.60} &\underline{11.08}\\
			6 &TaEr &5 &\Checkmark  &U$^{2}$ &P &9.25 &4.65 &\underline{2.81} &76.47 &11.03\\
			7 &TaEr &6 &\Checkmark  &U$^{2}$ &P &10.66 &4.79 &\textbf{2.83} &\textbf{76.84} &\textbf{11.15}\\
			\hline
			\blue{8} &TaErLite &0 &\Checkmark  &U &P &\textbf{0.18} &\textbf{0.07} &2.35 &61.07 &7.09\\
			\blue{9} &TaErLite &1 &\Checkmark   &U &P &\underline{0.87} &\underline{0.14} &2.51 &66.87 &8.48\\
			\blue{10} &TaErLite &2 &\Checkmark  &U &P &1.56 &0.21 &2.59 &69.38 &9.17\\
			\blue{11} &TaErLite &3 &\Checkmark  &U &P &2.26 &0.28 &2.63 &70.60 &9.56\\
			\blue{12} &TaErLite &4 &\Checkmark  &U &P &2.95 &0.35 &2.64 &70.89 &9.58\\
			\blue{13} &TaErLite &5 &\Checkmark  &U &P &3.64 &0.42 &2.67 &71.75 &9.71\\
			\blue{14} &TaErLite &6 &\Checkmark  &U &P &4.34 &0.49 &2.62 &70.93 &9.56\\
			\hline
			\blue{15} &TaEr &3 &\XSolidBrush  &U$^{2}$ &P &7.16 &6.01 &2.78 &76.21 &10.98\\
			\hline
			\blue{16} &TaEr &3 &\Checkmark  &U &P &5.13 &1.43 &2.75 &74.51 &10.56\\
			\blue{17} &TaEr &3 &\Checkmark  &Transformer &P &14.08 &1.42 &2.64 &71.35 &9.67\\
			\blue{18} &TaEr &3 &\Checkmark  &Conformer &P &29.93 &3.02 &2.70 &73.58 &10.37\\
			\hline
			\blue{19} &TaEr &3 &\Checkmark  &U$^{2}$ &E &6.42 &4.36 &2.77 &75.92 &10.82\\
			\bottomrule
	\end{tabular}}
	\vspace{-0.3cm}
\end{table*}
\begin{table}[!t]
	\renewcommand\arraystretch{1.20}
	\large
	\centering
	\caption{\blue{$p$-value comparisons among different expansion orders for TaEr.}}
	\label{tbl:p-values-taer}
	\resizebox{0.92\columnwidth}{!}{
		\begin{tabular}{c|ccccccc}
			\hline
			\blue{Order} &\blue{0} &\blue{1} &\blue{2} &\blue{3} &\blue{4} &\blue{5} &\blue{6}\\
			\hline
			\blue{0} &\blue{n.a} &\blue{$<$0.05} &\blue{$<$0.05} &\blue{$<$0.05} &\blue{$<$0.05} &\blue{$<$0.05} &\blue{$<$0.05}\\
			\blue{1} &\blue{-} &\blue{n.a} &\blue{$<$0.05} &\blue{$<$0.05} &\blue{$<$0.05} &\blue{$<$0.05} &\blue{$<$0.05}\\
			\blue{2} &\blue{-} &\blue{-} &\blue{n.a} &\blue{0.3585} &\blue{0.1252} &\blue{0.2237} &\blue{$<$0.05}\\
			\blue{3} &\blue{-} &\blue{-} &\blue{-} &\blue{n.a} &\blue{0.2163} &\blue{0.3455} &\blue{0.1070}\\
			\blue{4} &\blue{-} &\blue{-} &\blue{-} &\blue{-} &\blue{n.a} &\blue{0.6496} &\blue{0.3244}\\
			\blue{5} &\blue{-} &\blue{-} &\blue{-} &\blue{-} &\blue{-} &\blue{n.a} &\blue{0.2007}\\
			\blue{6} &\blue{-} &\blue{-} &\blue{-} &\blue{-} &\blue{-} &\blue{-} &\blue{n.a}\\
			\hline
		\end{tabular}}
	\vspace{-0.2cm}
	\end{table}
\section{Experimental Setup}
\label{sec:experimental-setup}
\subsection{Dataset Preparation}
\label{sec:datasets}
For single-channel setting, three datasets are adopted, namely WSJ0-SI84~{\cite{paul1992design}}, DNS-Challenge dataset~{\cite{reddy2020interspeech}}, and Voice-bank+Demand~{\cite{valentini2016investigating}}. \blue{Two datasets are utilized for multi-channel setting, namely spatialized Librispeech~{\cite{panayotov2015librispeech}} and L3DAS22 multi-channel speech enhancement challenge datasets{\cite{Guizzo}}. Detailed illustrations towards these datasets are given below.}
\begin{itemize}
	\item \textbf{WSJ0-SI84}: It consists of 7138 clean utterances by 83 speakers (41 females and 42 males). 77 speakers are selected, where 5428 and 957 utterances are split for training and validation, respectively. The remaining 6 unseen speakers are for model evaluation. To synthesize noisy-clean training pairs, we randomly select 20,000 types of noises from the DNS-Challenge noise set, whose duration is around 55 hours. Following~{\cite{li2021two}}, the training SNR is sampled from [-5, 0] dB with an interval of 1dB. In total, we generate around 150,000 and 10,000 noisy-clean pairs as the training and validation sets, respectively, and the total duration is around 300 hours. For model testing, two cases are set, namely Set-A, and Set-B. In Set-A, three challenging non-stationary noises are chosen, namely babble and factory taken from NOISEX92 database~{\cite{varga1993assessment}}, and cafeteria taken from CHiME3~{\cite{barker2015third}}. To evaluate the performance in more general noisy environments, in Set-B, around 50 environmental noises from MUSAN noise set~{\cite{snyder2015musan}} are adopted. For both cases, three SNRs are chosen, namely \{-5, 0, 5\} dB, and 400 pairs are set for each noise and each SNR.
	\item \textbf{DNS-Challenge}: It provides over 500 hours of clean clips spoken by 2150 speakers, and over 180 hours of noise clips for training. Besides, it provides a non-blind validation set for model comparison, which includes 150 noisy-clean pairs. Based on the scripts provided by the organization{\footnote{https://github.com/microsoft/DNS-Challenge}}, we totally generate around 3000 hours of pairs as the training set, in which the SNR ranges from -5 dB to 15 dB.
	\item \textbf{Voicebank+Demand (VB)}: There are around 30 speakers, where 28 are used in the training set and the remaining 2 speakers are for testing. For the training set, \blue{40 types of noise are mixed with clean speech to generate 11,572 noisy-clean pairs under four SNRs}, \emph{i.e.}, \{0, 5, 15, 20\} dB, and 20 noises are mixed with clean speech to yield 824 pairs under four SNRs, \emph{i.e.}, \{2.5, 7.5, 12.5, 17.5\} dB.
	\item \textbf{Spatialized Librispeech}: We use the open-sourced LibriSpeech corpus to synthesize the multi-channel dataset, where \textit{train-clean-100}, \textit{dev-clean}, and \textit{test-clean} are chosen for training, validation, and testing, respectively. For directional noises, we also use the noises taken from DNS-Challenge noise set. Similar to~{\cite{tan2022neural}}, two different microphone arrays are employed, namely a circular array of seven microphones and a uniform linear array (ULA) of seven microphones. For the circular array, one microphone is placed in the center and the remaining six microphones are uniformly placed in a circle. The radius of the circle is 4.25 cm. For the ULA, the distance between adjacent microphones is 5 cm. The room size is randomly chosen from 5-5-3 m to 10-10-4 m (length-width-height), and the reverberation time (RT$_{60}$) is randomly selected from 100 ms to 1000 ms, where the first 100 ms of the room impulse response (RIR) is convolved with the clean speech to obtain the target speech. For each target speech, we place 1-3 directional noises in the space and the distance between the noise and microphone is randomly taken from 0.5 m to 5.0 m. The training SNR is in the range of [-5, 10] dB. For testing, 1-3 directional noises sampled from the MUSAN noise set are spatially randomly placed with the SNR range of [-5, 5] dB and 200 pairs are generated. Within each utterance, the positions of all sources are assumed to be static.
	
	\item \blue{\textbf{L3DAS22 Challenge}: It is required to reconstruct the original dry speech source signal from a far-field noisy-reverberant office environments. The clean speech sounds are taken from the Librispeech dataset and background noises are from FSD50K{\cite{fsd50k}}. The SNR is sampled from the range of [6, 16] dBFS (decibels relative to full scale). The simulation uses real RIRs recorded in an office room using two first-order A-format Ambisonic arrays, each with four microphones. The microphone placement is fixed, with one Ambisonic microphone array placed at the room center and the other 20 cm away. The room configuration is the same for both training and testing, and the source positions are uniformly sampled inside the room with no overlap of positions between the two sets. Noisy mixtures are created by convolving the dry speech and noise signals with the measured RIRs first, and the convolved signals are then added together. There are 37,398, 2,362, and 2,189 mixture-clean pairs in the training, validation, and test sets, respectively. The generated A-format Ambisonic mixtures are converted to B-format Ambisonic via a transformation consisting of a pre-filter, a mixing matrix, and a post-filter.}
\end{itemize}
\begin{table}[!t]
	\renewcommand\arraystretch{1.20}
	\large
	\centering
	\caption{\blue{$p$-value comparisons among different expansion orders for TaErLite.}}
	\label{tbl:p-values-taerlite}
	\resizebox{0.92\columnwidth}{!}{
		\begin{tabular}{c|ccccccc}
			\hline
			\blue{Order} &\blue{0} &\blue{1} &\blue{2} &\blue{3} &\blue{4} &\blue{5} &\blue{6}\\
			\hline
			\blue{0} &\blue{n.a} &\blue{$<$0.05} &\blue{$<$0.05} &\blue{$<$0.05} &\blue{$<$0.05} &\blue{$<$0.05} &\blue{$<$0.05}\\
			\blue{1} &\blue{-} &\blue{n.a} &\blue{$<$0.05} &\blue{$<$0.05} &\blue{$<$0.05} &\blue{$<$0.05} &\blue{$<$0.05}\\
			\blue{2} &\blue{-} &\blue{-} &\blue{n.a} &\blue{$<$0.05} &\blue{$<$0.05} &\blue{$<$0.05} &\blue{$<$0.05}\\
			\blue{3} &\blue{-} &\blue{-} &\blue{-} &\blue{n.a} &\blue{0.4429} &\blue{0.1228} &\blue{0.4796}\\
			\blue{4} &\blue{-} &\blue{-} &\blue{-} &\blue{-} &\blue{n.a} &\blue{0.1566} &\blue{0.5365}\\
			\blue{5} &\blue{-} &\blue{-} &\blue{-} &\blue{-} &\blue{-} &\blue{n.a} &\blue{0.8644}\\
			\blue{6} &\blue{-} &\blue{-} &\blue{-} &\blue{-} &\blue{-} &\blue{-} &\blue{n.a}\\
			\hline
	\end{tabular}}
\vspace{-0.2cm}
\end{table}
\subsection{Miscellaneous Configurations}
\label{sec:detailed-configurations}
\vspace{-0.05cm}
\subsubsection{Parameter configurations for TaEr}
\label{sec:parameter-configurations-taer}
For the proposed TaEr, in the 0th-order, the kernel size and stride of the 2D-(De)GLU are $\left(1, 3\right)$ and $\left(1, 2\right)$ along the time and frequency axes, respectively, and $\left(2, 3\right)$ and $\left(1, 2\right)$ for UNet-block. The number of channels for all the 2D-Convs remains 64 by default. For encoder, the depth of the UNet-block $U_{en} = \left\{4, 3, 2, 1, 0\right\}$, and $U_{de} = \left\{1, 2, 3, 4, 0\right\}$ for decoder, where $0$ means that no UNet-block is used. For bottleneck, two S-TCN groups are utilized, in each of which four S-TCMs are stacked with dilation rates being \{1, 2, 5, 9\}. In each S-TCM, the kernel size and channel number are \{1, 256\} for input and output 1D-Convs, and \{5, 64\} for dilated 1D-Convs. In each high-order module, two groups of S-TCN are adopted, and the size of the LSTM is 256. The number of high-order terms $Q$ is empirically set as 3 and we also investigate the impact of $Q$ in the ablation study.
\subsubsection{Parameter configurations for TaErLite}
\label{sec:parameter-configurations-taerlite}
In the 0th-order module, the frequency size after the ERB filter-bank is reduced to 32, and the size of the GRU layers for magnitude filtering is 128 with group size of 2. In the high-order encoder, kernel size and the number of 2D-Convs are $\left(1, 3\right)$ and 32, respectively. In each high-order module, the size of the GRU layers is 256 with a group size of 2. For the auxiliary post-filter, the size of GRU layers is set to 32.
\subsubsection{Training configurations}
\label{sec:training-configurations}
All the utterances are sampled at 16 kHz. 20 ms square-root Hann window is used as both analysis and synthesis windows and 50\% overlap is adopted between adjacent frames. \blue{Following our preliminary work~{\cite{li2022glance}}, 320-point FFT is adopted, leading to 161 frequency bands.} As the final estimation involves the superimposition of the intermediate 0th-order and high-order terms, similar to~{\cite{li2022filtering}}, the supervision is only applied to the final result, \emph{i.e.}, both the magnitude filtering and residual estimation are adaptively learned by the network, which endows the network with more optimization freedom. The RI loss with the magnitude constraint is adopted, and the complex power spectrum compression strategy is adopted to decrease the dynamic range~{\cite{li2021importance}}. Adam optimizer~{\cite{kingma2014adam}} is adopted and we train the network for 60 epochs with a batch size of 6 in the utterance level. The learning rate is initialized at 5e-4 and will halve if the loss does not decrease after two consecutive epochs. To promote reproducibility, the source code will be available at \url{https://github.com/Andong-Li-speech/TaEr}.
\renewcommand\arraystretch{1.30}
\begin{table*}[t]
	\caption{\blue{Objective result comparisons among different models in terms of PESQ, eSTOI, SI-SNR, and overall DNS-MOS with ITU-T Rec. P.835 for Set-A and Set-B on the WSJ0-SI84 dataset.} ``Do.'' denotes the processing domain of the method, and time domain and time-frequency domain methods are briefly denoted by ``T'' and ``T-F'', respectively. ``Avg.'' refers to the average score among different SNRs and is specified in ``mean$\pm$std'' format.}
	\centering
	\huge
	\resizebox{\textwidth}{!}{
		\begin{tabular}{ccccccccccccccccccccc}
			\toprule
			&\multirow{2}*{Models}  &\multirow{2}*{Year} &\multirow{2}*{\rotatebox{90}{Do.}} &\multirow{2}*{\rotatebox{90}{Causal}}
			&\multicolumn{4}{c}{PESQ $\uparrow$} &\multicolumn{4}{c}{eSTOI (\%) $\uparrow$} &\multicolumn{4}{c}{SI-SNR (dB) $\uparrow$} &\multicolumn{4}{c}{DNS-MOS OVLR$\uparrow$}\\
			\cmidrule(lr){6-9}\cmidrule(lr){10-13}\cmidrule(lr){14-17}\cmidrule(lr){18-21}
			& & &&  &-5dB &0dB &5dB &\multicolumn{1}{c}{Avg.} &-5dB &0dB &5dB &\multicolumn{1}{c}{Avg.} &-5dB  &0dB &5dB &\multicolumn{1}{c}{Avg.} &-5dB  &0dB &5dB &\multicolumn{1}{c}{Avg.}\\
			\cline{1-21}
			\multirow{13}*{\rotatebox{90}{Set-A}}
			&\multicolumn{1}{|c|}{Noisy} &- &- &- &1.54 &1.86 &2.17 &1.85$\pm$0.37 &29.25 &43.11 &57.53 &43.30$\pm$13.28 &-5.00 &0.00 &5.00 &0.00$\pm$4.09 &1.25 &1.59 &2.00 &1.61$\pm$0.48\\
			&\multicolumn{1}{|c|}{DDAEC} &2020 &T &\Checkmark &2.27 &2.79 &3.16 &2.74$\pm$0.45 &63.12 &76.65 &84.73 &74.83$\pm$11.34 &7.22 &11.23 &14.15 &10.87$\pm$3.42 &2.61 &2.86 &2.98 &2.82$\pm$0.27\\
			&\multicolumn{1}{|c|}{DEMUCAS} &2020 &T &\Checkmark &2.27 &2.73 &3.09 &2.69$\pm$0.49 &64.90 &77.72 &85.61 &76.08$\pm$10.82 &7.61 &11.50 &14.62 &11.24$\pm$3.35 &\underline{2.65} &\underline{2.91} &\underline{3.05} &\underline{2.87$\pm$0.26}\\
			&\multicolumn{1}{|c|}{GCRN} &2020 &T-F &\Checkmark &2.02 &2.55 &\blue{2.92} &\blue{2.50$\pm$0.45} &56.44 &72.83 &\blue{82.08} &\blue{70.45$\pm$13.13} &5.36 &9.72 &\blue{12.67} &\blue{9.25$\pm$3.70} &2.18 &2.58 &2.83 &2.53$\pm$0.36\\
			&\multicolumn{1}{|c|}{DCCRN} &2020 &T-F &\Checkmark &1.90 &2.46 &2.84 &2.40$\pm$0.48 &50.98 &68.06 &78.73 &65.92$\pm$14.06 &4.17 &8.61 &11.74 &8.17$\pm$3.83 &2.12 &2.34 &2.56 &2.34$\pm$0.30\\
			&\multicolumn{1}{|c|}{FullSubNet} &2021 &T-F &\Checkmark &2.20 &2.64 &2.97 &2.60$\pm$0.38 &50.44 &67.34 &78.88 &65.56$\pm$13.63 &4.34 &9.01 &12.81 &8.72$\pm$3.90 &2.24 &2.50 &2.74 &2.49$\pm$0.30\\
			&\multicolumn{1}{|c|}{CTSNet} &2021 &T-F &\Checkmark &2.32 &2.79 &3.14 &2.75$\pm$0.40 &62.92 &76.20 &84.35 &74.49$\pm$11.12 &6.75 &10.84 &13.96 &10.52$\pm$3.48 &2.41 &2.72 &2.91 &2.68$\pm$0.30\\
			&\multicolumn{1}{|c|}{GaGNet} &2022 &T-F &\Checkmark &2.32 &2.82 &3.19 &2.77$\pm$0.44 &64.93 &77.46 &85.25 &75.88$\pm$10.63 &7.23 &11.10 &14.11 &10.81$\pm$3.31 &2.49 &2.76 &2.94 &2.73$\pm$0.30\\
			&\multicolumn{1}{|c|}{FRNet} &2022 &T-F &\Checkmark &2.36 &2.85 &3.22 &2.81$\pm$0.44 &65.84 &78.13 &85.79 &76.59$\pm$10.47 &7.36 &11.23 &14.31 &10.97$\pm$3.33 &2.53 &2.79 &2.97 &2.76$\pm$0.29\\
			&\multicolumn{1}{|c|}{NCA} &2022 &T\&T-F &\Checkmark &\textbf{2.56} &\textbf{3.06} &\textbf{3.34} &\textbf{2.99$\pm$0.43} &\textbf{70.06} &\textbf{81.48} &\textbf{87.89} &\textbf{79.81$\pm$9.77} &\textbf{8.26} &\textbf{12.26} &\textbf{15.35} &\textbf{11.96$\pm$3.43} &\textbf{2.87} &\textbf{3.02} &\textbf{3.08} &\textbf{2.99$\pm$0.20}\\
			\rowcolor[HTML]{E7E6E6}
			&\multicolumn{1}{|c|}{TaErLite} &2022 &T-F &\Checkmark &2.25 &2.72 &3.08 &2.68$\pm$0.41 &59.23 &73.54 &82.60 &71.79$\pm$11.72 &6.18 &10.12 &13.16 &9.82$\pm$3.25 &2.32 &2.63 &2.86 &2.60$\pm$0.32\\
			\rowcolor[HTML]{E7E6E6}
			&\multicolumn{1}{|c|}{TaEr} &2022 &T-F &\Checkmark &\underline{2.42} &\underline{2.91} &\underline{3.28} &\underline{2.87$\pm$0.43} &\underline{67.33} &\underline{79.47} &\underline{86.57} &\underline{77.76$\pm$10.21} &\underline{7.75} &\underline{11.63} &\underline{14.73} &\underline{11.35$\pm$3.37} &2.61 &2.84 &2.99 &2.81$\pm$0.27\\
			\hline
			\multirow{13}*{\rotatebox{90}{Set-B}}
			&\multicolumn{1}{|c|}{Noisy} &- &- &- &1.74 &2.04 &2.41 &2.06$\pm$0.51 &44.59 &57.38 &69.45 &57.14$\pm$17.44 &-5.00 &0.00 &5.00 &0.00$\pm$4.08 &1.65 &2.05 &2.43 &2.04$\pm$0.57\\
			&\multicolumn{1}{|c|}{DDAEC} &2020 &T &\Checkmark &2.83 &3.17 &3.43 &3.15$\pm$0.43 &75.57 &83.65 &89.03 &82.75$\pm$10.26 &10.91 &13.67 &16.16 &13.58$\pm$3.63 &2.82 &2.93 &3.02 &2.92$\pm$0.22\\
			&\multicolumn{1}{|c|}{DEMUCAS} &2020 &T &\Checkmark &2.76 &3.09 &3.35 &3.07$\pm$0.44 &76.88 &84.56 &89.65 &83.69$\pm$9.70 &11.22 &14.03 &16.67 &13.97$\pm$3.86 &\underline{2.88} &\underline{3.01} &\underline{3.08} &\underline{2.99$\pm$0.20}\\
			&\multicolumn{1}{|c|}{GCRN} &2020 &T-F &\Checkmark &2.55 &2.94 &3.21 &2.90$\pm$0.49 &70.31 &80.82 &86.91 &79.35$\pm$12.26 &9.01 &12.27 &14.69 &11.99$\pm$4.04 &2.52 &2.78 &2.94 &2.75$\pm$0.32\\
			&\multicolumn{1}{|c|}{DCCRN} &2020 &T-F &\Checkmark &2.47 &2.89 &3.17 &2.84$\pm$0.55 &66.43 &77.92 &84.66 &76.34$\pm$13.69 &8.40 &11.74 &13.82 &11.32$\pm$4.10 &2.39 &2.53 &2.64 &2.52$\pm$0.28\\
			&\multicolumn{1}{|c|}{FullSubNet} &2021 &T-F &\Checkmark &2.66 &3.00 &3.30 &2.99$\pm$0.47 &67.08 &78.20 &85.81 &77.03$\pm$13.12 &8.87 &12.30 &15.60 &12.26$\pm$4.49 &2.54 &2.74 &2.89 &2.72$\pm$0.28\\
			&\multicolumn{1}{|c|}{CTSNet} &2021 &T-F  &\Checkmark &2.86 &3.19 &3.45 &3.17$\pm$0.45 &76.60 &84.12 &89.28 &83.33$\pm$10.12 &10.99 &13.84 &16.44 &13.75$\pm$4.06 &2.73 &2.89 &3.00 &2.87$\pm$0.25\\
			&\multicolumn{1}{|c|}{GaGNet} &2022 &T-F &\Checkmark  &2.91 &3.25 &3.52 &3.23$\pm$0.47 &77.60 &84.87 &89.82 &84.10$\pm$9.68 &11.11 &13.88 &16.49 &13.83$\pm$4.01 &2.77 &2.92 &3.01 &2.90$\pm$0.24\\
			&\multicolumn{1}{|c|}{FRNet} &2022 &T-F &\Checkmark &2.94 &3.28 &3.56 &3.26$\pm$0.46 &78.24 &85.30 &90.10 &84.55$\pm$9.48 &11.23 &14.00 &16.56 &13.93$\pm$3.94 &2.79 &2.93 &3.02 &2.92$\pm$0.24\\
			&\multicolumn{1}{|c|}{NCA} &2022 &T\&T-F &\Checkmark &\textbf{3.11} &\textbf{3.43} &\textbf{3.65} &\textbf{3.40$\pm$0.41} &\textbf{81.31} &\textbf{87.48} &\textbf{91.53} &\textbf{86.77$\pm$8.43} &\textbf{12.17} &\textbf{14.94} &\textbf{17.63} &\textbf{14.92$\pm$4.15} &\textbf{2.99} &\textbf{3.05} &\textbf{3.09} &\textbf{3.04$\pm$0.17}\\
			\rowcolor[HTML]{E7E6E6}
			&\multicolumn{1}{|c|}{TaErLite} &2022 &T-F &\Checkmark &2.76 &3.10 &3.38 &3.08$\pm$0.44 &73.33 &81.88 &87.66 &80.96$\pm$10.73 &9.76 &12.57 &15.01 &12.45$\pm$3.49 &2.65 &2.84 &2.96 &2.81$\pm$0.27\\
			\rowcolor[HTML]{E7E6E6}
			&\multicolumn{1}{|c|}{TaEr} &2022 &T-F &\Checkmark &\underline{3.01} &\underline{3.34} &\underline{3.60} &\underline{3.31$\pm$0.44} &\underline{79.70} &\underline{86.25} &\underline{90.86} &\underline{85.60$\pm$9.16} &\underline{11.63} &\underline{14.43} &\underline{17.10} &\underline{14.37$\pm$4.10} &2.83 &2.97 &3.05 &2.95$\pm$0.22\\
			\bottomrule

	\end{tabular}}
	\label{tbl:objective-results}
	\vspace{-0.2cm}
\end{table*}
\renewcommand\arraystretch{0.95}
\begin{table*}[t]
	\caption{\blue{Comparisons in terms of model size (in million) and computational complexity (in Giga) among models.}}
	\centering
	\small
	\resizebox{0.98\textwidth}{!}{
		\begin{tabular}{ccccccccccccc}
			\toprule
			Systems &DDAEC &DEMUCAS &GCRN &DCCRN &FullSubNet &CTSNet &GaGNet &FRNet &NCA &TaErLite &TaEr\\
			Param\# (M) &5.00 &18.87 &9.77 &\underline{3.67} &5.64 &4.44 &6.01 &7.52 &11.70 &\textbf{2.26} &6.42\\
			MACs (G/s) &36.85 &4.32 &2.40 &14.06 &29.83 &5.48 &\underline{1.64} &2.81 &6.37 &\textbf{0.28} &4.36\\
			RTF &0.97 &\underline{0.09} &\textbf{0.07} &0.28 &0.44 &0.19 &0.10 &0.19 &0.24 &0.15 &0.20\\
			\bottomrule
	\end{tabular}}
	\label{tbl:params-macs}
	\vspace{-0.2cm}
\end{table*}
\section{Results and Analysis}
\label{sec:results-and-analysis}
\vspace{-0.05cm}
\subsection{Ablation Study}
\label{sec:ablation-study}
For convenience, around 100 hours of the training set are sampled from WSJ0-SI84 corpus to conduct the ablation study, \blue{which spans the following four aspects: (1) the number of the expansion order $Q$; (2) whether the extracted feature $\mathbf{R}$ from the high-order encoder comes from the reused feature of the 0th-order module or not; (3) network type used in the 0th-order module and high-order encoder; (4) intermediate activation function type. All the involved models in the ablation study adopt the same training configurations.} Three objective metrics are chosen for this study, namely perceptual evaluation of speech quality (PESQ)~{\cite{rix2001perceptual}}, extended short-time objective intelligibility (eSTOI)~{\cite{jensen2016algorithm}}, and scale-invariant SNR (SI-SNR)~{\cite{le2019sdr}}. PESQ is to evaluate the speech quality with the range of $\left(-0.5, 4.5\right)$, eSTOI can measure the speech intelligibility, whose value ranges from 0 to 1. SI-SNR measures the level of speech distortion of the waveform and is not sensitive to the absolute energy level changes. For three metrics, a larger score indicates better performance. Besides, the number of trainable parameters and multiply-accumulate operations (MACs) are also provided as reference. Quantitative results are shown in Table~{\ref{tbl:ablation-study}}.
\begin{figure}[t]
	\centering
	\centerline{\includegraphics[width=0.85\columnwidth]{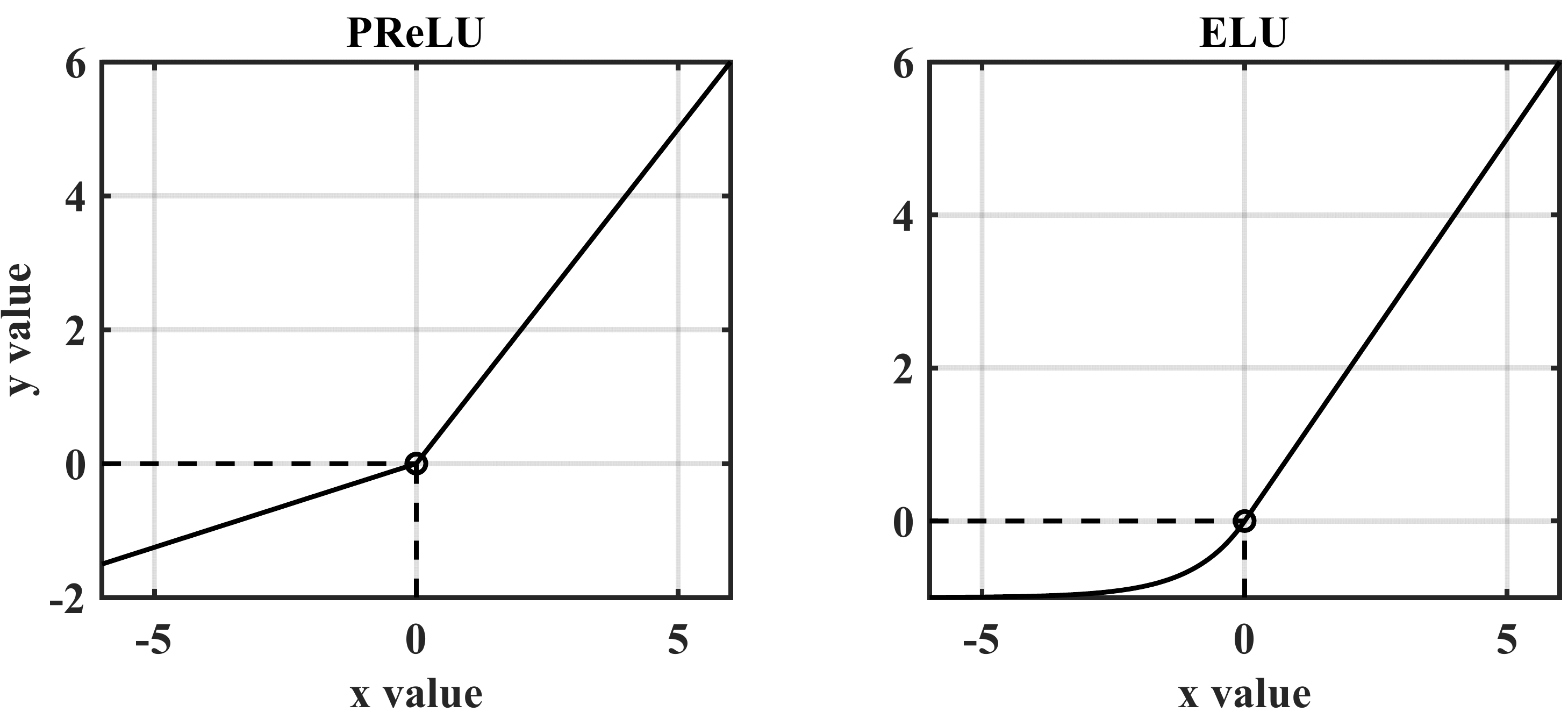}}
	\caption{Comparison between PReLU and ELU. PReLU is discontinuous at $X = 0$ while ELU is continuous everywhere.}
	\label{fig:prelu-elu}
	\vspace{-0.3cm}
\end{figure}
\subsubsection{Effect of the order number} 
For both TaEr and TaErLite, we explore the impact of Taylor's expansion order $Q$, which empirically enumerates from 0 to 6. When $Q = 0$, it is not surprising to observe the worst performance, which is because only the spectral magnitude is estimated and the noisy phase remains unaltered. With the increase of $Q$, we observe the notable performance improvements in three metrics. For example, from ID1 to ID2, around 0.12/4.63\%/1.70 dB score improvements are obtained in terms of PESQ/eSTOI/SI-SNR, respectively. Besides, when $Q$ further increases, \emph{e.g.}, $Q\ge 3$, only mild improvements and even slight degradations are obtained, and \blue{we attribute the main reason to that high-order terms aim for phase modification and spectral refinement, which usually exhibits a rather sparse distribution (see Fig.~{\ref{fig:example}}(d)). Thus, several high-order terms can be adequate for target modeling.} A similar trend is observed for TaErLite, as shown from \blue{ID8 to ID14}.

\blue{We perform one-tailed two-paired $t$-tests to assess the significance of various expansion orders. The null hypothesis is rejected if the $p$-value is less than 0.05, indicating that a higher value of $Q$ can result in a significant performance gain over a smaller value of $Q$. Tables {\ref{tbl:p-values-taer}} and {\ref{tbl:p-values-taerlite}} present the $p$-values for TaEr and TaErLite in terms of SI-SNR, respectively. The results indicate that when the 0th-order term is implemented using a complicated and powerful network structure, such as the encoder-decoder structure in TaEr, $Q=2$ may be sufficient for target modeling, and a higher value does not significantly improve performance. However, when a light-weight network is used for the 0th-order term, such as GRUs in the ERB domain for TaErLite, a larger $Q$ can still yield a significant metric improvement.}
\subsubsection{Effect of the high-order encoder shared scheme}
As the design of a separate high-order encoder often leads to larger parameter and computation overhead, we thus explore for TaEr, whether the encoded feature in the 0th-order module can be reused and fed to the latter high-order terms. It is worth emphasizing that for TaErLite, as the 0th-order is conducted in the ERB domain, where only the coarse-grained magnitude is adopted and the phase is dropped, a separate high-order encoder is imperative to extract the phase information. \blue{Comparing ID15 with ID4, after the feature reusing}, we do not observe the performance gain, e.g., PESQ: 2.78 vs. 2.78, eSTOI: 76.32\% vs. 76.21\%. We conjecture that the latter high-order modules are tasked with complex residual estimation, which is often sparse in the spectral sense, and the fine-granularity feature from the 0th-order module is thus adequate to involve the modeling of the residual component.
\subsubsection{Effect of the network type in the 0th-order module}
Except for U$^{2}$Net, we also do experiments on other network candidates in the 0th-order module, including U-Net, transformer, and conformer, as shown in \blue{ID16-ID18}. For U-Net, we drop all the UNet-blocks in REL and RDL. For transformer, six transformer layers are stacked, in each of which the feature dimension is set to 512 with 8 heads in parallel. \blue{Similarly, for conformer, six conformer layers are adopted}, in each of which the feature dimension, head number, and kernel size are set to $\left\{512, 8, 31\right\}$, respectively. \blue{From ID16 to ID4}, we observe notable improvements in three metrics, indicating the significance of UNet-block in fined-grained feature aggregation. For \blue{ID17 and ID18}, despite much more parameters, the performance is still much inferior to that of the proposed module. The reason might be although the transformer and conformer are adept at effectively establishing global sequence correlations, they often neglect the local patterns, especially in the harmonic regions, which are indispensable for spectrum recovery. In contrast, the adopted module incorporates both local convolutions and long-range TCNs and facilitates the estimation from a complementary perspective.
\subsubsection{Effect of the activation function}
Recall that we assume the function $\mathcal{G}\left(\cdot\right)$ to be differentiable to each order. However, we use the PReLU as the default activation function, which is not smooth at $X = 0$ (see Fig.~{\ref{fig:prelu-elu}(a)). To explore whether the discontinuous point will hamper the performance, we replace PReLU with the exponential linear unit (ELU)~{\cite{clevert2015fast}}, which is continuous and differentiable at all points, as shown in Fig.~{\ref{fig:prelu-elu}}(b). \blue{The results are shown in \blue{ID4 and ID19}.} One can see that for the ELU case, it is overall inferior to that of PReLU. We intend to emphasize that although the derivative operator may not hold for discontinuous points, instead of pursuing the rigorous consistency in the mathematical definition, we actually intend to mimic the behavior of Taylor's series approximation by replacing all the complicated and unstable operators with trainable and stable modules. Therefore, our method is more general and can be robust to discontinuous points.  
\subsection{Comparisons with Baselines on WSJ0-SI84 Corpus}
\label{sec:comparison-with-baselines-on-wsj0}
\vspace{-0.05cm}
Based on the abovementioned analysis, TaEr and TaErLite with the configurations of \blue{ID4 and ID11} respectively are adopted to compare with existing state-of-the-art (SOTA) approaches as they well balance between trainable parameters and computational complexity. \blue{Nine baselines are selected, namely DDAEC~{\cite{pandey2020densely}}, DEMUCAS~{\cite{defossez2020real}}, GCRN~{\cite{tan2019learning}}, DCCRN~{\cite{hu2020dccrn}}, FullSubNet~{\cite{hao2021fullsubnet}}, CTSNet~{\cite{li2021two}}, GaGNet~{\cite{li2022glance}}, FRNet~{\cite{li2022filtering}}, and NCA~{\cite{wang2021neural}}.} All the models are reimplemented with causal settings, including causal convolutions, uni-directional RNNs, and causal normalizations. Except for PESQ, eSTOI, and SI-SNR, DNS-MOS~{\cite{naderi2021crowdsourcing}}, a more recently proposed metric, which follows ITU-T Rec. P.835 and is highly related with the subjective rating of humans, is also adopted. We only list the overall quality score (OVRL) herein. Quantitative results are shown in Table~{\ref{tbl:objective-results}}, and several conclusions can be drawn. (1) For both test sets, all the models can notably improve the performance in terms of four metrics, indicating the effectiveness of existing NN-based SE methods in handling both highly non-stationary and more general environmental noises. (2) When comparing between the time domain and T-F domain models, overall, the latter still dominates the performance superiority. For example, although GCRN and DCCRN are relatively inferior to ConvTasNet and DPRNN, CTSNet and its successors can break the performance limit by following the target decoupling paradigm. Moreover, by incorporating time and T-F prior knowledge, NCA, a hybrid-domain model, further improve the performance. (3) \blue{Overall, the proposed TaEr surpasses the previous top-performing methods and is only inferior to NCA}, which can be explained by two factors. First, although NCA is causally implemented in the first and third stages, the second stage is operated in the time domain with a frame length of 128 ms and thus can utilize more contextual information, while that of 20 ms for our method. Besides, the combination of time and T-F domain can introduce complementary prior over single-domain methods. (4) It is interesting to observe that despite the parameters, computation complexity and buffer length are dramatically reduced, the proposed TaErLite still achieves comparable or even better performance over some baselines. It fully reveals the superiority of the proposed Taylor's approximation paradigm.
\renewcommand\arraystretch{1.00}
\begin{table}[!t]
	\caption{\blue{Comparisons in terms of trainable parameters (in kilo), computational complexity (in Mega), and time receptive field size between TaEr and TaErLite.}}
	\centering
	\resizebox{0.99\columnwidth}{!}{
		\begin{tabular}{ccccc}
			\toprule
			&\multicolumn{2}{c}{\blue{0th-order}}  &\multicolumn{2}{c}{\blue{high-order}}\\
			\cmidrule(lr){2-3}\cmidrule(lr){4-5}
			 &\blue{TaEr} &\blue{TaErLite} &\blue{TaEr} &\blue{TaErLite}\\
			\hline
			\blue{Param\# (K)}  &\blue{2174} &\blue{110 (94.9\%$\downarrow$)} &\blue{1414} &\blue{693 (51.0\%$\downarrow$)} \\
			\blue{MACs (M/s)}  &\blue{3929} &\blue{11 (99.7\%$\downarrow$)} &\blue{143} &\blue{70 (51.0\%$\downarrow$)} \\
			\blue{Receptive Field}  &\blue{177} &\blue{2 (98.9\%$\downarrow$)} &\blue{137} &\blue{2 (98.5\%$\downarrow$)} \\
			\bottomrule
	\end{tabular}}
	\label{tbl:taer-taerlite}
	\vspace{-0.2cm}
\end{table}

The number of trainable parameters, MACs per second, and real time factor (RTF) of different models are shown in Table~{\ref{tbl:params-macs}}, where the RTF value is averaged by repeating the measurements for five times with an Intel Core (TM) i7-9700 CPU clocked at 3.00 GHz. One can get that overall, the proposed TaEr obtains decent overhead and processing speed among the baselines. For TaErLite, we observe the smallest value in terms of trainable parameters and computation complexity. \blue{Table~\ref{tbl:taer-taerlite} presents the number of trainable parameters, computational complexity, and receptive field size of the 0th-order and high-order terms for TaEr and TaErLite, where $Q$ is set to 1 as an example. As demonstrated, for the 0th-order module, replacing the U-Net structure with the light-weight grouped GRU layer under ERB scale in TaErLite results in a significant reduction in the number of parameters and MACs by 94.9\% and 99.7\%, respectively. Moreover, replacing all the S-TCMs with GRUs leads to a notable decrease in the receptive field size, from 177 to 2 in the 0th-order and from 137 to 2 in the high-order term. These results validate the prominent advantages of TaErLite in edge computing devices.}
\renewcommand\arraystretch{1.00}
\begin{table}[!t]
	\caption{Comparisons on the DNS-Challenge non-blind test set.}
	\huge
	\centering
	\resizebox{0.99\columnwidth}{!}{
		\begin{tabular}{cccccc}
			\hline
			Models &Year &WB-PESQ &PESQ &STOI (\%) &SI-SNR (dB)\\
			\hline
			Noisy  &- &1.58 &2.45 &91.52 &9.07\\
			NSNet~\cite{reddy2020interspeech}  &2020 &2.15 &2.87  &94.47 &15.61\\
			DTLN~{\cite{westhausen2020dual}}  &2020 &- &3.04 &94.76  &16.34\\
			DCCRN~{\cite{hu2020dccrn}}  &2020 &- &3.27 &- &- \\
			PoCoNet~{\cite{isik2020poconet}} &2020 &2.75 &- &- &-\\
			FullSubNet~{\cite{hao2021fullsubnet}} &2021 &2.78  &3.31 &96.11 &17.29\\
			TRU-Net~{\cite{choi2021real}} &2021 &2.86  &3.36 &96.32 &17.55\\
			DCCRN+~{\cite{lv2021dccrn+}} &2021 &- &3.33 &- &- \\
			CTS-Net~{\cite{li2021two}} &2021 &2.94 &3.42  &96.66  &17.99\\
			GaGNet~{\cite{li2022glance}} &2022 &3.17 &\underline{3.56} &97.13 &18.91\\
			FRNet~{\cite{li2022filtering}}  &2022  &3.14 &3.52 &96.91 &18.75\\
			MDNet~{\cite{li2022mdnet}}  &2022 &3.18 &\underline{3.56} &97.20 &19.17\\
			FullSubNet+~{\cite{chen2022fullsubnet+}} &2022 &2.98 &3.50 &96.69 &18.34\\
			FS-CANet~{\cite{chen22k}} &2022 &3.02  &3.51  &96.74 &18.08\\
			FRCRN{\cite{zhao2022frcrn}}      &2022  &\underline{3.23}  &\textbf{3.60} &\textbf{97.69} &\textbf{19.78}\\
			STSubNet~{\cite{xiong2022spectro}}   &2022 &3.00 &- &97.03  &\underline{19.64}\\
			\rowcolor[HTML]{E7E6E6}
			TaErLite  &2022 &2.78 &3.33 &95.96 &17.18\\
			\rowcolor[HTML]{E7E6E6}
			TaEr   &2022 &\textbf{3.26}  &\textbf{3.60} &\underline{97.56} &19.40\\
			\hline
	\end{tabular}}
	\label{tbl:dns}
	\vspace{-0.2cm}
\end{table}
\renewcommand\arraystretch{1.00}
\begin{table}[!t]
	\caption{Comparisons on the Voicebank+Demand dataset.}
	\centering
	\large
	\resizebox{0.99\columnwidth}{!}{
		\begin{tabular}{cccccc}
			\toprule
			\multicolumn{1}{c}{Model} & \multicolumn{1}{c}{Year} & \multicolumn{1}{c}{WB-PESQ}  & \multicolumn{1}{c}{CSIG} & \multicolumn{1}{c}{CBAK} &
			\multicolumn{1}{c}{COVL}\\
			\midrule
			Noisy &- &1.97 &3.35 &2.44 &2.63\\
			SEGAN~{\cite{pascual2017segan}} &2017 &2.16 &3.48 &2.94 &2.80 \\
			MMSE-GAN~{\cite{soni2018time}} &2018 &2.53 &3.80 &3.12 &3.14 \\
			CRGAN~{\cite{zhang2020loss}}  &2020 &2.92 &4.16 &3.24 &3.54\\
			MetricGAN~{\cite{fu2019metricgan}} &2019 &2.86 &3.99 &3.18 &3.42 \\
			DCCRN~{\cite{hu2020dccrn}}  &2020 &2.68 & 3.88 & 3.18 & 3.27\\
			PHASEN~{\cite{yin2020phasen}} &2020 &2.99 &4.21 &3.55 &3.62 \\
			T-GSA~{\cite{kim2020t}} &2020 &3.06 &4.18 &\underline{3.59} &3.62\\
			CTSNet~{\cite{li2021two}} &2021 &2.92 &4.25 &3.46 &3.59\\
			MetricGAN+~{\cite{fu2021metricgan+}}  &2021 &\underline{3.15} &4.14 & 3.16 & 3.64\\
			SE-Conformer~{\cite{eesung2021se}}  &2021 &3.13 &\underline{4.45} & 3.55 &\underline{3.82}\\
			GaGNet~{\cite{li2022glance}}  &2022 &2.94 & 4.26 & 3.45 & 3.59 \\
			FRNet~{\cite{li2022filtering}} &2022 &3.06 &4.32 &3.53 &3.68 \\
			DBTNet~{\cite{yu2022dbt}}  &2022 &\textbf{3.30} &\textbf{4.59}  &\textbf{3.75} &\textbf{3.92}\\
			\rowcolor[HTML]{E7E6E6}
			TaErLite &2022 &2.78 &4.10 &3.40 &3.44\\
			\rowcolor[HTML]{E7E6E6}
			TaEr  &2022 &3.07 &4.35  &3.58 &3.73\\
			\bottomrule
	\end{tabular}}
	\label{tbl:vb-result}
	\vspace{-0.4cm}
\end{table}
\renewcommand\arraystretch{1.00}
\begin{table*}[t]
	\caption{\blue{Objective result comparisons among different models in terms of PESQ, eSTOI, SI-SNR, and overall DNS-MOS with ITU-T Rec. P.835 on the spatialized Librispeech  dataset.} Two microphone arrays are adopted, namely circular array and linear array with seven microphones for each array. The models are separately trained for each case. The best and second best results for NN-based methods are \textbf{boldfaced} and \underline{underlined}, respectively.}
	\centering
	\Huge
	\resizebox{\textwidth}{!}{
		\begin{tabular}{ccccccccccccc}
			\toprule
			\multirow{2}*{Models} &\multirow{2}*{Year} &\multirow{2}*{\rotatebox{90}{Causal}} &Param\# &MACs  &\multicolumn{4}{c}{Circular Array (7-channels)} &\multicolumn{4}{c}{Linear Array (7-channels)}\\
			\cmidrule(lr){6-9}\cmidrule(lr){10-13}
			& & &(M) &(G/s) &PESQ $\uparrow$&eSTOI (\%) $\uparrow$ &SI-SNR (dB)$\uparrow$ &DNS-MOS OVLR $\uparrow$  &PESQ $\uparrow$&eSTOI (\%) $\uparrow$ &SI-SNR (dB)$\uparrow$ &DNS-MOS OVLR $\uparrow$\\
			\cline{1-13}
			Nosiy &- &- &- &- &1.63 &41.13 &-1.89 &1.11 &1.60 &41.12 &-1.85 &1.15\\
			MMUB~{\cite{9596418}} &2021 &\Checkmark &\textbf{1.96} &8.35 &2.26 &60.86 &\underline{5.66} &2.22 &2.17 &60.86 &\underline{5.34} &2.29\\
			NSF~{\cite{tan2022neural}} &2022 &\Checkmark &16.24 &5.33 &2.37 &63.34 &4.62 &2.28 &2.37 &65.01 &4.40 &\underline{2.33}\\
			COSPA~{\cite{9747528}} &2022 &\Checkmark &3.21 &1.13 &2.31 &61.07 &5.29 &2.03 &2.23 &57.54 &4.07 &2.00\\
			EaBNet~{\cite{li2022embedding}} &2022 &\Checkmark &2.83 &7.45 &\textbf{2.79} &\textbf{73.78} &\textbf{7.53} &\underline{2.30} &\textbf{2.69} &\textbf{72.64} &\textbf{7.07} &2.32\\
			\rowcolor[HTML]{E7E6E6}
			TaErLite &2022 &\Checkmark &\underline{2.27} &\textbf{0.31} &2.33 &61.16 &4.12 &2.13 &2.24 &59.78 &3.60 &2.08\\ 
			\rowcolor[HTML]{E7E6E6}
			TaEr &2022 &\Checkmark &6.45 &4.44 &\underline{2.57} &\underline{67.74} &5.50 &\textbf{2.35} &\underline{2.50} &\underline{67.00} &5.19 &\textbf{2.35}\\
			(Oracle) Frame-MVDR &- &\Checkmark &- &- &2.46 &72.38 &6.35 &1.73 &2.45 &71.80 &4.88 &1.81\\
			\hline
			FasNet-TAC (4ms)~{\cite{luo2020end}} &2020 &\XSolidBrush &\underline{2.65} &16.52 &2.59 &71.51 &\underline{8.69} &2.40 &2.38 &65.50 &6.59 &2.32\\
			NSF~{\cite{tan2022neural}} &2022 &\XSolidBrush &12.96 &4.99 &2.63 &70.12 &5.43 &\underline{2.63} &2.58 &68.80 &5.02 &\underline{2.63}\\
			COSPA~{\cite{9747528}} &2022 &\XSolidBrush &3.66 &\underline{1.16} &2.40 &63.92 &6.30 &2.08 &2.31 &61.26 &5.19 &2.08\\
			FT-JNF~{\cite{9944916}} &2022 &\XSolidBrush &3.35 &54.36 &2.81 &72.93 &7.58 &2.27 &2.70 &70.13 &6.77 &2.19\\
			EaBNet~{\cite{li2022embedding}} &2022 &\XSolidBrush &2.82 &7.44 &\textbf{3.04} &\textbf{79.69} &\textbf{8.82} &2.56 &\textbf{2.94} &\textbf{77.98} &\textbf{8.06} &2.61\\
			\rowcolor[HTML]{E7E6E6}
			TaErLite &2022 &\XSolidBrush &\textbf{1.96} &\textbf{0.28} &2.50 &65.80 &5.03 &2.42 &2.51 &66.71 &4.97 &2.44\\ 
			\rowcolor[HTML]{E7E6E6}
			TaEr &2022 &\XSolidBrush &6.03 &4.38 &\underline{2.84} &\underline{74.10} &6.68 &\textbf{2.66} &\underline{2.89} &\underline{75.91} &\underline{6.78} &\textbf{2.69}\\
			(Oracle) TI-MVDR &- &\XSolidBrush &- &- &2.46 &72.39 &7.91 &1.82 &2.46 &72.39 &7.91 &1.79\\
			(Oracle) TI-MWF &- &\XSolidBrush &- &- &2.56 &75.01 &10.79 &1.77 &2.51 &73.65 &9.76 &1.79\\
			\bottomrule
	\end{tabular}}
	\label{tbl:objective-mc}
	\vspace{-0.2cm}
\end{table*}
\subsection{Comparisons with Baselines on DNS-Challenge and Voicebank+Demand benchmark}
\label{sec:comparison-with-baselines-on-dns-vb}
\vspace{-0.05cm}
\blue{Quantitative comparisons with previous top-performing baselines on DNS-Challenge and VB benchmarks are presented in Tables~{\ref{tbl:dns}}-{\ref{tbl:vb-result}}.} For DNS-Challenge, four objective metrics are adopted, namely, the wideband PESQ (abbreviated as WB-PESQ)~{\cite{itu862}}, PESQ, STOI~{\cite{taal2010short}}, and SI-SNR. For VB dataset, WB-PESQ and another three MOS-related objective metrics~{\cite{hu2007evaluation}} are utilized, namely CSIG, CBAK, and COVL. Note that as most of the previous baselines in the VB are adopted in the offline format, we thus convert TaEr and TaErLite into the non-causal setting. From the tables, one can get that on the DNS-Challenge dataset, \blue{TaEr notably surpasses other top-performing baselines and is comparable with FRCRN~{\cite{zhao2022frcrn}},} a most recently proposed SOTA model. On the VB corpus, TaEr does not achieve the best performance, and we attribute the reason to the vanilla network module design, which is expected to improve in the future. Besides, our lite version, TaErLite, also achieves modest scores and is competitive with existing methods. \blue{Therefore, the efficacy superiority of the proposed methods can be revealed.}
\begin{table}[t]
	\caption{\blue{Comparisons on the L3DAS22 Speech Enhancement Task.}}
	\centering
	\Huge
	\resizebox{0.99\columnwidth}{!}{
		\begin{tabular}{cccc}
			\toprule
			\multicolumn{1}{c}{\blue{Models}} &\multicolumn{1}{c}{\blue{WER (\%) $\downarrow$}} &\multicolumn{1}{c}{\blue{STOI (\%) $\uparrow$}}  &\blue{Task-Metric (\%) $\uparrow$}\\
			\midrule
			\blue{Baseline (MMUB)} &\blue{21.2} &\blue{87.8} &\blue{83.3}\\
			\blue{ESP-SE} &\blue{\textbf{1.89}} &\blue{\textbf{98.7}} &\blue{\textbf{98.4}}\\
			\blue{BaiduSpeech} &\blue{2.50} &\blue{97.5} &\blue{97.5}\\
			\blue{PCG-AIID}  &\blue{3.20} &\blue{97.2} &\blue{97.0}\\
			\rowcolor[HTML]{E7E6E6}
			\blue{TaEr} &\blue{7.66}  &\blue{93.1}  &\blue{92.7}\\
			\bottomrule
	\end{tabular}}
	\label{tbl:l3das22-challenge}
	\vspace{-0.4cm}
\end{table}
\begin{figure*}[t]
	\centering
	\centerline{\includegraphics[width=0.99\textwidth]{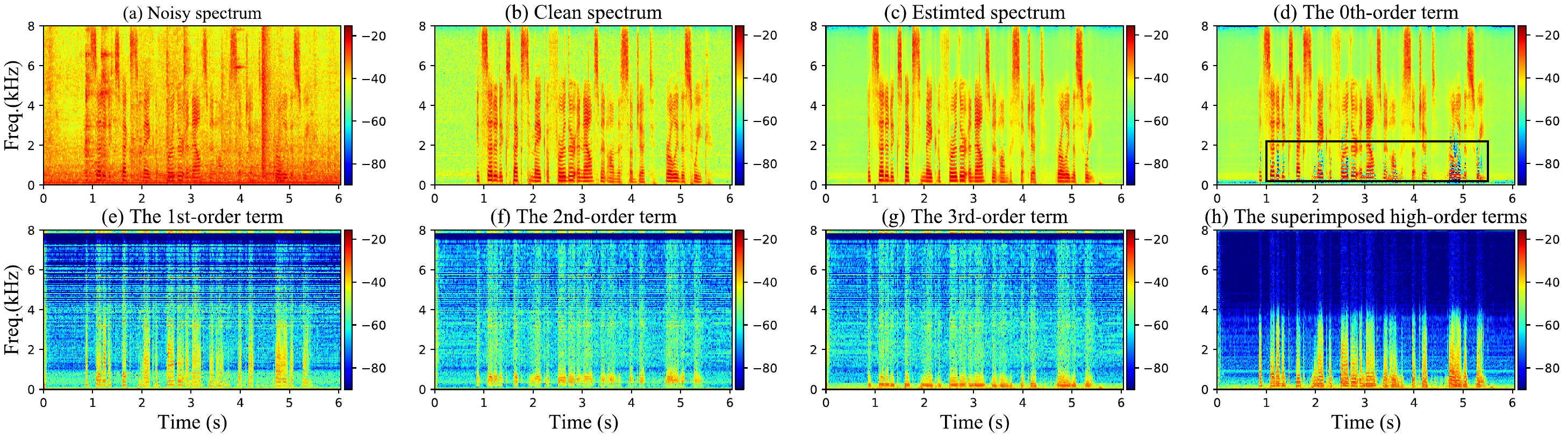}}
	\caption{Spectral visualization of the estimated 0th-order term and high-order terms when $Q = 3$. Speech file name is ``01jc020e.wav''.}
	\label{fig:visualization-order-terms}
	\vspace{-0.3cm}
\end{figure*}
\begin{figure}[t]
	\centering
	\subfigure[\blue{TaEr}]{
		\includegraphics[width= 0.46\columnwidth]{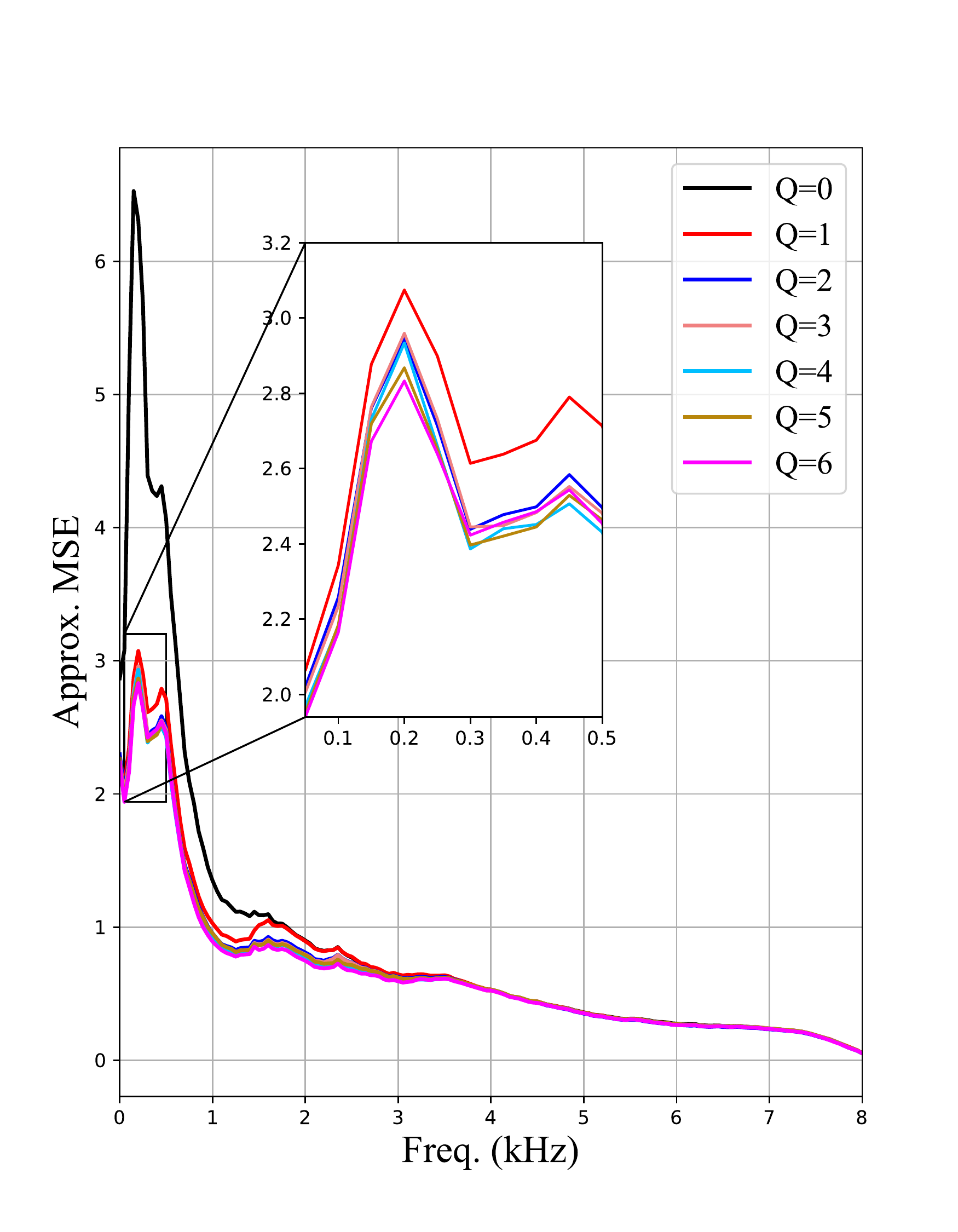}	
	}
	\subfigure[\blue{TaErLite}]{
		\includegraphics[width= 0.46\columnwidth]{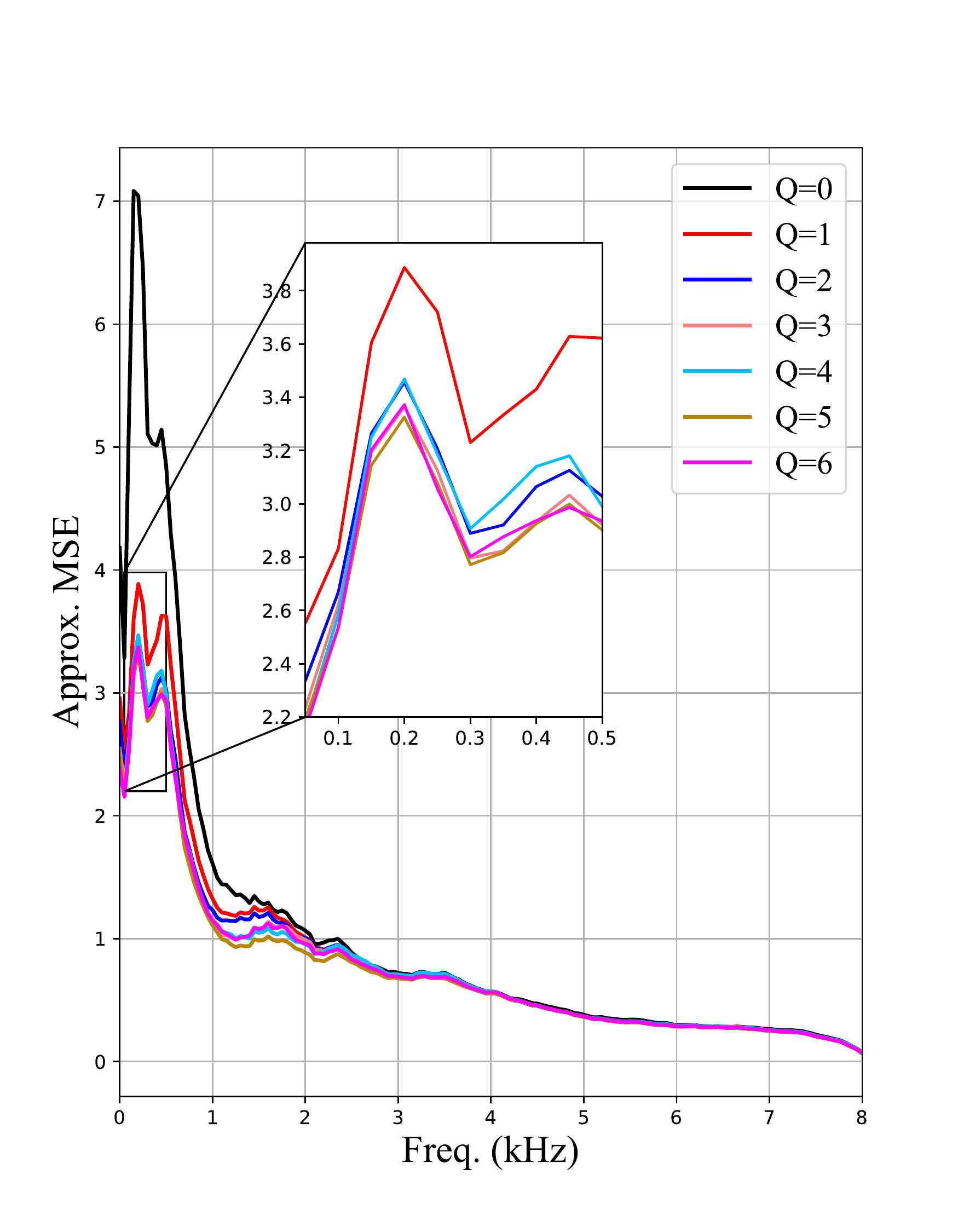}	
	}
	\caption{\blue{Approximation MSE between the network estimation and target speech under different number of expansion orders. (a) TaEr. (b) TaErLite.}}
	\label{fig:approximation-mse}
	\vspace{-0.2cm}
\end{figure}
\subsection{Comparisons with Baselines for Multi-Channel Case}
\label{sec:comparison-with-baselines-on-multi-channel-case}
\vspace{-0.05cm}
As stated in Section~{\ref{sec:proposed-taylor-unfolding-framework}}, the proposed model can easily adapt to the multi-channel setting by concatenating RI spectra of different microphones as the network input. Table~{\ref{tbl:objective-mc}} presents quantitative comparisons with advanced multi-channel methods on the spatialized Librispeech dataset under both causal and non-causal settings. Note that except for NN-based models, \blue{we also present the results of the frame-level MVDR, time-invariant MVDR (TI-MVDR), and TI-MWF}, where the speech and noise covariance matrices are calculated with oracle speech and noise. From the table, several observations can be made. First, the non-causal setting yields better performance over the causal setting, \emph{e.g.}, for TaEr, 0.27/6.36\%/1.18 dB/0.31 gains are achieved in PESQ/eSTOI/SI-SNR/OVLR for circular array, revealing the significance of future information in speech recovery. Second, overall, the proposed TaEr surpasses other NN baselines and is inferior to EaBNet, where the beamforming weights of each channel are predicted frame-by-frame. The reason can be explained as in EaBNet, the spatial information can be better utilized by adopting frame-level beamforming. In contrast, we extend our method into the multi-channel version by regarding the signals from different microphones as more input information sources of the general function to estimate the magnitude and complex residual of the reference channel, and therefore, the spatial cue may not be fully utilized, which needs to be further investigated as future research. In addition, despite oracle traditional beamformers (TI-MVDR, TI-MWF) are not advantageous in PESQ, they yield relatively better performance in SI-SNR, which can be explained as we assume the positions of all sources to be static within each utterance, and traditional beamformers can preserve the target speech with low distortion via linear spatial filtering under oracle condition. Therefore, it still remains to be a non-trivial problem how to decrease the speech distortion for existing neural beamformers especially in extreamely low SNR scenarios.

\blue{Table~{\ref{tbl:l3das22-challenge}} presents quantitative results of the proposed TaEr and other SOTA baselines on the L3DAS Challenge dataset. Three objective metrics are used: word error rate (WER), STOI, and their combination metric defined as}
\begin{equation}
\label{eqn30}
\blue{\text{Task-Metric} = \left(\text{STOI} + \left(\text{1} - \text{WER}\right)\right) / 2.}
\end{equation}

\blue{Four baselines are considered, where MMUB serves as the challenge baseline, and ESP-SE~{\cite{luyj}}, BaiduSpeech~{\cite{guochang}}, and PCG-AIID~{\cite{jingdong}} are the top three schemes. The results show that TaEr outperforms the challenge baseline in all three metrics but works worse than the SOTA methods. To further improve the performance, two strategies can be considered. First, as shown in~{\cite{jingdong}}, introducing the filter-and-sum operation can notably improve performance over implicitly estimating the target speech. Second, as noted by the authors in~{\cite{luyj}}, the introduction of a linear spatial filter, \emph{e.g.}, multi-frame MWF, can effectively reduce speech distortion, improving recognition performance.}
\renewcommand\arraystretch{0.82}
\begin{table}[t]
	\caption{\blue{MSE of the actual remaining error term for different number of orders. All the values are averaged upon Set-A of the WSJ0-SI84 dataset.}}
	\centering
	\tiny
	\resizebox{0.50\columnwidth}{!}{
		\begin{tabular}{ccc}
			\toprule
			\blue{$Q$} &\multicolumn{1}{c}{\blue{TaEr}} &\multicolumn{1}{c}{\blue{TaErLite}}\\
			\midrule
			\blue{0} &\blue{0.29} &\blue{0.23}\\
			\blue{1} &\blue{0.12} &\blue{0.11}\\
			\blue{2} &\blue{0.11} &\blue{0.10}\\
			\blue{3}  &\blue{0.11} &\blue{0.10}\\
			\blue{4} &\blue{0.10}  &\blue{0.11}\\
			\blue{5} &\blue{0.11}  &\blue{0.12}\\
			\bottomrule
	\end{tabular}}
	\label{tbl:error-blund}
	\vspace{-0.4cm}
\end{table}
\subsection{\blue{Visualization of Each Order Term}}
\label{sec:visualization-of-each-order-term}
\vspace{-0.05cm}
In this section, we investigate the role of the 0th-order and high-order terms. Spectral visualizations of the estimation of each order are shown in Fig.~{\ref{fig:visualization-order-terms}}. One can see that the 0th-order is already close to the clean target in the spectral distribution, indicating that most of the interferences have been removed with the 0th-order approximation. Besides, the overall distributions from the 1st to the 3rd order module seem quite sparse and we can notice the contour in the harmonic regions, indicating that high-order terms are mainly adopted to capture the residual component rather than for dense prediction. An interesting observation is that in the filtered spectrum after the 0th-order module, we find some valleys in the neighboring harmonics, indicating that some extra spectral energy may be leaked to the high-order terms, as shown in the black box area in Fig.~{\ref{fig:visualization-order-terms}}(d). This is because we do not provide extra supervision toward the output of the 0th-order module in the training process, that is to say, the network actually attempts to allocate the role of the magnitude filter and complex-residual estimator automatically~{\cite{li2022filtering}}. Besides, as all the modules are jointly trained without any intermediate constraint, we cannot guarantee strict consistency between the theoretical distribution of each order term in Taylor's theorem and their actual network outputs. Nonetheless, from the visualization, the 0th-order and high-order terms still behave as a coarse magnitude filter and complex-residual generator, which is consistent with our expectations.
\subsection{\blue{Model Convergence Analysis}}
\label{sec:model-convergence-analysis}
\blue{In Taylor's series expansion, the error bound property often needs to be considered. Let us denote the superimposed spectrum estimation as $\left\{\mathcal{T}_{r}\left(Q, X\right), \mathcal{T}_{i}\left(Q, X\right)\right\}$:}
\begin{equation}
\label{eqn31}
\blue{\mathcal{T}_{r}\left(Q, X\right) = \sum_{q=0}^{Q}\frac{1}{q!}\mathcal{H}_{r}\left(q, X, \delta_{r}\right),}
\end{equation}
\begin{equation}
\label{eqn32}
\blue{\mathcal{T}_{i}\left(Q, X\right) = \sum_{q=0}^{Q}\frac{1}{q!}\mathcal{H}_{i}\left(q, X, \delta_{i}\right).}
\end{equation}

\blue{Ideally, the Taylor's approximation will become more accurate with the increase of $Q$, \emph{i.e.},}
\begin{equation}
\label{eqn33}
\blue{\mathcal{T}_{r}\left(Q+1, X\right) = \mathcal{T}_{r}\left(Q, X\right) + \frac{1}{\left(Q+1\right)!}\mathcal{H}_{r}\left(Q+1, X, \delta_{r}\right),}
\end{equation}
\begin{equation}
\label{eqn34}
\blue{\mathcal{T}_{i}\left(Q+1, X\right) = \mathcal{T}_{i}\left(Q, X\right) + \frac{1}{\left(Q+1\right)!}\mathcal{H}_{i}\left(Q+1, X, \delta_{i}\right),}
\end{equation}
\blue{where the real and imaginary parts of the remaining term $\left\{\frac{1}{\left(Q+1\right)!}\mathcal{H}_{r}\left(Q+1, X, \delta_{r}\right), \frac{1}{\left(Q+1\right)!}\mathcal{H}_{i}\left(Q+1, X, \delta_{i}\right)\right\}$ tend to converge to a smaller value and cannot exceed an error bound. As the proposed method simulates the decomposition and superimposition processes of Taylor's theorem, we calculate the MSE of the actual remaining error when the number of orders is set to $Q$ and $Q+1$, respectively, as shown in Table~{\ref{tbl:error-blund}}:}
\begin{equation}
\label{eqn35}
\begin{split}
\blue{\text{MSE}\left(Q, X\right)} &\blue{= \frac{1}{KL}\sum_{l}\sum_{k}\left|\mathcal{T}_{r}\left(Q+1, X_{l, k}\right) - \mathcal{T}_{r}\left(Q, X_{l, k}\right)\right|^{2}} \\
& \blue{+ \frac{1}{KL}\sum_{l}\sum_{k}\left|\mathcal{T}_{i}\left(Q+1, X_{l, k}\right) - \mathcal{T}_{i}\left(Q, X_{l, k}\right)\right|^{2}}
\end{split}
\end{equation}
\blue{where $Q\in\left\{0, 1, 2, 3, 4, 5\right\}$. Note that the network is trained separately for each $Q$ value. From the table, it can be observed that for both TaEr and TaErLite, $Q = 0$ yields a larger MSE value over other cases. However, the MSE value does not converge to a smaller value from $Q = 2$ to $Q = 5$. We attribute the reason to the adopted end-to-end training strategy, where the estimation of each order term is automatically allocated by the network. This means that we cannot guarantee identical intermediate results among different $Q$ settings. Therefore, the actual remaining term not only includes $\frac{1}{\left(Q+1\right)!}\mathcal{H}_{r/i}\left(Q+1, X, \delta_{r}\right)$, but also involves the intrinsic gap in the lower order terms, say, $\frac{1}{q!}\mathcal{H}_{r/i}^{'}\left(q, X, \delta_{r/i}\right) - \frac{1}{q!}\mathcal{H}_{r/i}^{''}\left(q, X, \delta_{r/i}\right)$ and $q\in\left\{0,\cdots,Q\right\}$, which may prevent the actual remaining term from further decreasing. Here, the superscripts $\left(\cdot\right)^{'}$ and $\left(\cdot\right)^{''}$ are used to distinguish the actual estimation under different $Q$ values. To better analyze the error bound, an alternative option is to estimate each order term sequentially, which is regarded as future work. For instance, we can consider only the 0th-order term, train it until network convergence, freeze its parameters, and then train the 1st-order term until its convergence, and so on. By doing so, the lower terms between different $Q$ values can be identical, and only the highest order term needs to be considered.}

\blue{In Fig.~{\ref{fig:approximation-mse}}, we present the MSE values between the superimposed spectrum estimation and target with different $Q$ values, which are calculated upon Set-A of the WSJ0-SI84 dataset. A notable MSE decrease is observed from $Q = 0$ to $Q = 1$, especially under 1 kHz frequency regions, indicating that high order terms can profit the target approximation. When $Q$ value further increases, \emph{e.g.}, from 1 to 5, the approximation MSE will gradually converge, which is due to the limited mapping capability of the network.}
\section{Concluding Remarks}
\label{sec:conclusion}
In this paper, we propose a general all-neural unfolding framework for single-channel and multi-channel speech enhancement tasks based on Taylor's theorem. Specifically, we revisit the complex spectrum recovery problem and formulate it into the magnitude mapping in the neighborhood space of the noisy mixture, \blue{where the complex residual term is introduced for phase refinement in advance}. Based on that, we decompose the non-linear mapping function into the superimposition of the 0th-order and high-order terms, where the former aims to provide a coarse magnitude estimation while the latter is designed for residual estimation in the complex spectrum domain. All the terms are instantiated with trainable parameters to warrant end-to-end training. We carry out comprehensive experiments on both simulated  single-channel and multi-channel datasets, and \blue{quantitative results show that the proposed model achieves competitive performance with existing state-of-the-art SE systems and also enjoys better flexibility.}

As our work is orthogonal to existing fundamental architecture design, we hope to employ more advanced network structures in the future, so as to further push the upper-bound performance. Besides, as the method in essence follows a general progressive approximation learning paradigm, it has the potential to address other speech signal processing problems, \emph{e.g.}, speech dereverberation, acoustic echo cancellation, and audio/speech source separation, in which existing NN-based methods still simply follow the empirical guidance and may lack adequate flexibility to deploy in the practical devices. Finally, we do not impose any intermediate constraint in the training process, which may cause the de-facto estimations to be inconsistent with the mathematical definitions. Therefore, it remains to be solved how to effectively control the output of each order so that the internal interpretability can be further strengthened. 
%
\bibliographystyle{IEEEbib}

\end{document}